\documentclass[conference,a4paper,10pt]{IEEEtran}
\usepackage{amsmath,amssymb}
\usepackage{amsthm}
\usepackage{psfrag}
\usepackage{epsfig}
\usepackage{graphicx,subfigure}
\usepackage{mathrsfs}
\usepackage{url}
\usepackage{bbm,color}
\usepackage[compress]{cite}
\usepackage{mathtools}
\usepackage{array}
\usepackage{supertabular}
\newtheorem{theorem}{Theorem}

\newtheorem{remark}{Remark}

\begin{document}
\title{Kinematic Resolutions of Redundant Robot Manipulators using Integration-Enhanced RNNs}

\author{Lingdong Kong\\
\\
School of Automation Science and Engineering,\\
South China University of Technology\\
}

\maketitle

\begin{abstract}
  Recently, a time-varying quadratic programming (QP) framework that describes the tracking operations of redundant robot manipulators is introduced to handle the kinematic resolutions of many robot control tasks. Based on the generalization of such a time-varying QP framework, two schemes, i.e., the Repetitive Motion Scheme and the Hybrid Torque Scheme, are proposed. However, measurement noises are unavoidable when a redundant robot manipulator is executing a tracking task. To solve this problem, a novel integration-enhanced recurrent neural network (IE-RNN) is proposed in this paper. Associating with the aforementioned two schemes, the tracking task can be accurately completed by IE-RNN. Both theoretical analyses and simulations results prove that the residual errors of IE-RNN can converge to zero under different kinds of measurement noises. Moreover, practical experiments are elaborately made to verify the excellent convergence and strong robustness properties of the proposed IE-RNN.
\end{abstract}

\section{Introduction}\label{sec.introduction}

The inverse kinematics of robot manipulators can be described as a series of mathematical processes, which makes use of formulas and transformations to determine joint parameters like joint-velocities and joint-accelerations \cite{Antonelli2009Stability,NYR2018A,zhang2020modification}. Such parameters offer a desired position for each robot manipulator' end-effector. There are two traditional strategies for solving inverse kinematics problems, i.e., the closed-form solution and the numerical solution \cite{Craig1986Introduction}. The closed-form solution, which is based on analytic expression, includes two approaches, i.e., geometric and algebraic schemes \cite{Abdel2012A,Shimizu2008Analytical}. Due to the iterative feature of numerical solution, the time cost of closed-form solution is usually shorter. Therefore, the closed-form solution is more popular in the early years. However, with the rapid development of technology and the high demand of human requirement, the mechanical structure of robot manipulators are designed to be more and more complex \cite{Merat2003Introduction}.
Under this case, the closed-form solution cannot be found \cite{Craig1986Introduction}. And how to design a fast and effective numerical algorithm has become the focus of current research \cite{Angeles1985On,Merat2003Introduction}.

The numbers of degrees-of-freedom (termed as DOF) of robot manipulators are increasing to adapt to complicated working environments. Generally, a five or six DOF manipulator can locate the position and orientation of end-effector in 3-D space \cite{Craig1986Introduction}. But when facing obstacles or joint-limits, some position/orientation might not be achieved \cite{Craig1986Introduction,Merat2003Introduction}. Therefore, redundant robot manipulators, whose kinematics problems cannot be solved by closed-form solution, are becoming more and more important in the robot industry. The traditional numerical method to solve the inverse kinematics of redundant robot manipulators is the pseudo-inverse method \cite{Chevallereau1987Efficient}.

One of the most serious problems of pseudo-inverse method is that it cannot solve the singularity and bound constrain of robot manipulators. In recent years, a quadratic programming (termed as QP) method is introduced to remedy this problem \cite{Toshani2014Real}. On the basis of the Lagrange theory and the Karush-Kuhn-Tucker condition, the aforementioned QP problem can be considered as a optimization problem \cite{Zhang2013Variable}, and further solved by some numerical methods. He \textit{et al.} translated this constrained optimization problem into an equivalent linear-projection equation, and proposed four numerical methods to solve this projection equation \cite{He1994Solving}. A QP-based motion generation scheme for dual arms of humanoid robot is proposed in Ref. \cite{KLD2018A}, and further converted into a linear variational inequality (termed as LVI). In Refs. \cite{Zhang2017Three,8665072}, three numerical methods, i.e., 94LVI, E47, and M4 methods are exploited for the online joint-free redundancy resolution of PUMA560 manipulator.

It is worth pointing out that the aforementioned numerical iterative methods are simple to derive, but hard to compute, and their convergence are only guaranteed for a limited class of matrices. Recently, with the development of neural network, some recurrent neural network (termed as RNN) based methods are proposed for solving inverse kinematics of robot manipulator \cite{Tejomurtula1999Inverse}. A QP-based neural method is proposed by Toshani \textit{et al.} for handling real-time inverse kinematics problems \cite{Toshani2014Real}. In Ref. \cite{Xia2000A}, Xia \textit{et al.} proposed a RNN method to solve linear projection equations. A cerebellum-inspired neural network is firstly introduced by Mitra \textit{et al.} for solving inverse kinematics problems \cite{Mitra2015Cerebellum}. In Ref. \cite{Qu2009Real}, a modified pulse-coupled neural network (termed as MPC-NN) is studied to achieve real-time robot path planning. To obtain the inverse resolution of wheeled mobile manipulator, a zeroing recurrent neural network (termed as Z-RNN) is proposed by Xiao \textit{et al.}. The global exponential convergence feature of Z-RNN is proved in Refs. \cite{Zhang2009Zhang,9097909}. In Ref. \cite{Zhang2010Robustness}, the robustness of Z-RNN is theoretically analyzed by Zhang \textit{et al.}, and a high dimensional situation is discussed and verified by MATLAB simulations. In Ref. \cite{Zhang2008The}, the links between Z-RNN and the Newton iteration method are discussed and compared. The Z-RNN method, which is activated by a li-function, is surveyed and applied to manipulation of robot by Guo \textit{et al.} \cite{Guo2014Li}.

Due to the parallel processing feature of RNN, the computation speed of RNN is faster than that of numerical methods \cite{9072323}. And by using the optical and very-large-scale integration (termed as VLSI) strategy, RNN can be easily realized by electronic circuit components \cite{8877851,8558699}. However, Z-RNN has a serious problem. When facing noise disturbance, which is a general phenomenon in practical application of robot manipulator, the robustness performance of Z-RNN could be deteriorated \cite{ZhangA2017}. This disadvantage could lead to task failure when robot manipulator is executing path tracking, goods capturing, electric welding, and other missions. More seriously, this may damage the mechanical manipulator itself or even the human body. Therefore, it is necessary for us to consider an anti-noise algorithm which can meet the needs of our desire.

Inspired by the design idea of Z-RNN, in this paper, an integration-enhanced recurrent neural network (termed as IE-RNN) is proposed to obtain the kinematics resolution of redundant robot manipulator. By respectively considering the inverse kinematics problems at joint-velocity and joint-acceleration level, two schemes, i.e., the Repetitive Motion Scheme and the Hybrid Torque Scheme are introduced and converted to a QP framework. On the basis of the Lyapunov stability \cite{Mazenc2006Further} and Nyquist's theory \cite{Luse1988A}, the convergence and robustness of IE-RNN for solving this QP problem are carefully analyzed. Different form the traditional Z-RNN method, IE-RMM has excellent performance when handling noises polluted kinematics problems. Both computer simulations and practical experiment results demonstrate the feasibility, accuracy, and superiority of IE-RNN for tracking complex pathes.

The reminder of this paper is organized as follows. In Section II, the Repetitive Motion Scheme and the Hybrid Torque Scheme are formulated and translated into a standard QP form. Section III gives the design process of the proposed neural model. The convergence and robustness of the neural model are theoretically analyzed in Section IV. Section V gives the computer simulations and practical experiments. Section VI draws the conclusion and discusses the future work.

The main contributions of this paper are listed as follows.
\begin{itemize}
  \item An integration-enhanced recurrent neural network (IE-RNN) is proposed for kinematics resolution of redundant robot manipulator.
  \item The convergence performance of IE-RNN with power-sigmoid activation function is theoretically proved.
  \item The robustness of IE-RNN perturbed by constant noise and time-varying noise is analyzed.
  \item Computer simulations and practical experiments further verify the effectiveness of IE-RNN.
\end{itemize}

\section{Problem Formulation} \label{sec.Problem Formulation}

Due to the redundancy and nonlinearity of a perturbed redundant robot manipulator, it is difficult to straightly consider and  obtain its unique solution, hence it is necessary for us to take the inverse kinematics into consideration at the velocity level. The relationship between the end-effector velocity and the joint velocity can be describe as the following form, i.e.,
\begin{eqnarray}
\mathcal{J}(\theta(t))\dot{\theta}(t)=\mathcal{\dot{R}}(t)
\label{eqn.FCRM}
\end{eqnarray}
where $\theta(t)\in\mathbb{R}^{n}$ is the joint space vector; $\mathcal{R}(t)\in\mathbb{R}^{n}$ is the desired end-effector path vector; and their derivatives $\dot{\theta}(t)$, $\mathcal{\dot{R}}(t)$ denote the joint angular velocity and the end-effector velocity, respectively; $\mathcal{J}(\theta)\in\mathbb{R}^{m\times n}$ is the Jacobian matrix which is defined as $\mathcal{J}(\theta)=\partial\mathcal{F}(\theta)/\partial\theta$ and $\mathcal{F}(\theta)$ denotes the forward-kinematics mapping.

\subsection{Repetitive Motion Scheme}

In order to obtain the solution of Equation (\ref{eqn.FCRM}), a QP-based formulation form is leaded to describe the feedback control and repetitive motion, i.e.,
\begin{eqnarray}
\begin{split}
\text{minimize~~}   &\frac{1}{2}||\mathcal{M}(t)+\dot{\theta}(t)||_{2}^{2}\\
\text{subject to~} &\mathcal{J}(\theta(t))\dot{\theta}(t)=\mathcal{\dot{R}}(t)+\mathcal{W}(\mathcal{R}(t)-\mathcal{F}(\theta))
\end{split}
\label{eqn.QP-RM}
\end{eqnarray}
where $\mathcal{M}(t)=\kappa(\theta(t)-\theta(0))$ with $\kappa>0$ is the magnitude of the response to the joint drift $\theta(t)-\theta(0)$; $\mathcal{W}(t)\in\mathbb{R}^{m\times m}$ denotes the feedback-controlled matrix; and $||\cdot||_{2}$ denotes the Euclidean norm of a vector.

\subsection{Hybrid Torque Scheme}

Consider the kinematics problem at the joint-acceleration level, a QP-based formulation is introduced to describe the input of the torque control with feedback, i.e.,
\begin{eqnarray}
\begin{split}
\text{minimize~~}  &\frac{1}{2}(\mu||\mathcal{M}+\dot{\theta}||_{2}^{2}+(1-\mu)||\mathcal{T}(\theta)||_{2}^{2})\\
\text{subject to~} &\mathcal{J}\ddot{\theta}=\mathcal{\ddot{R}}-\mathcal{\dot{J}}\dot{\theta}+\alpha(\mathcal{\dot{R}}-\mathcal{J}\theta)
+\beta(\mathcal{R}-\mathcal{F}(\theta))\\
&~\mathcal{T}(\theta)=\mathcal{I}(\theta)\ddot{\theta}+\mathcal{C}(\theta,\dot{\theta})+\mathcal{G}(\theta)
\end{split}
\label{eqn.QP-HT}
\end{eqnarray}
where $\mathcal{M}(t)$, $\mathcal{J}(t)$, $\theta(t)$, $\mathcal{R}(t)$ and their derivatives are defined exactly the same as before; $\mu\in[0,1]$ denotes the weight coefficient; $\mathcal{T}(\theta)\in\mathbb{R}^{n}$ denotes the joint torque vector; $\mathcal{I}(\theta)\in\mathbb{R}^{n\times n}$ denotes the inertia matrix; $\mathcal{C}(\theta,\dot{\theta})\in\mathbb{R}^{n}$ severally denotes the Coriolis force vector and centrifugal force vector; and $\mathcal{G}(\theta)\in\mathbb{R}^{n}$ denotes the gravitational force vector.

\subsection{Standard QP Form}

For the convenience of our following discussion and make it easier and more clear to be intelligible, the aforementioned two schemes (i.e. Repetitive Motion Scheme (\ref{eqn.QP-RM}) and Hybrid Torque Scheme (\ref{eqn.QP-HT})) can be further reformulated into a standard QP form, i.e.,
\begin{eqnarray}
\begin{split}
&\text{minimize~~~} \frac{1}{2}x^{\text{T}}(t)\mathcal{Q}(t)x(t)+\mathcal{P}^{\text{T}}(t)x(t)\\
&\text{subject to~~} \mathcal{J}(t)x(t)=\mathcal{B}(t)
\end{split}
\label{eqn.QP}
\end{eqnarray}
where $\mathcal{J}(\theta)\in\mathbb{R}^{m\times n}$ is remain the Jacobian matrix defined as above; superscript $^{\text{T}}$ denotes the transpose operation of a vector or a matrix.\\
$\bullet$ For Repetitive Motion Scheme (\ref{eqn.QP-RM}):\\ $x(t):=\dot{\theta}(t)\in\mathbb{R}^{n}$; $\mathcal{P}(t):=\mathcal{M}(t)=\kappa(\theta(t)-\theta(0))\in\mathbb{R}^{n}$ with $\kappa>0$; $\mathcal{B}(t):=\mathcal{\dot{R}}(t)+\mathcal{W}(\mathcal{R}(t)-\mathcal{F}(\theta))\in\mathbb{R}^{n}$; and $\mathcal{Q}(t):=\mathrm{I}(t)\in\mathbb{R}^{n\times n}$ denotes the identity matrix.\\
$\bullet$ For Hybrid Torque Scheme (\ref{eqn.QP-HT}):\\ $x(t):=\ddot{\theta}(t)\in\mathbb{R}^{n}$; $\mathcal{Q}(t):=\mu\mathrm{I}(t)+(1-\mu)\mathcal{I}^{2}(t)\in\mathbb{R}^{n\times n}$; $\delta:=(1-\mu)\mathcal{I}^{\text{T}}(t)(\mathcal{P}(t)+\mathcal{G}(t))+\mu\mathcal{S}(t)$ with $\mathcal{S}(t)=(\xi_{1}+\xi_{2})\dot{\theta}(t)+\xi_{1}\xi_{2}(\theta(t)-\theta(0))$; and $\mathcal{B}(t):=\mathcal{\ddot{R}}(t)-\mathcal{\dot{J}}(t)\dot{\theta}(t)+\alpha(\mathcal{\dot{R}}-\mathcal{J}\theta)
+\beta(\mathcal{R}-\mathcal{F}(\theta))\in\mathbb{R}^{n}$.

\section{Neural Model}
In this section, a novel integration-enhanced recurrent neural network (IE-RNN) is proposed. For comparison, the traditional neural method, i.e., zeroing recurrent neural network (Z-RNN) is also demonstrated.

The design process of IE-RNN contains the following five steps.

\textbf{Step 1:}
In order to obtain the solution of the standard time-varying QP problem (\ref{eqn.QP}), a Lagrange form of this problem is constructed as below
\begin{eqnarray}
\begin{split}
\mathcal{L}\big(x(t),\lambda(t),t\big)=&\frac{1}{2}x^{\text{T}}(t)\mathcal{Q}(t)x(t)+\mathcal{P}^{\text{T}}(t)x(t)+
\\&\lambda^{\text{T}}(t)\big(\mathcal{J}(t)x(t)-\mathcal{B}(t)\big), \ t\in [0,+\infty)
\label{eqn.Lagrange-1}\
\end{split}
\end{eqnarray}
where $\lambda(t)\in \mathbb{R}^m$ denotes the Lagrangian multiplier vector.

\textbf{Step 2:}
As for QP problem (\ref{eqn.QP}), according to the Lagrangian theory \cite{Pagilia2001Control}, if both $\partial\mathcal{L}\big(x(t),\lambda(t),t\big)/\partial x(t)$ and $\partial \mathcal{L}\big(x(t),\lambda(t),t\big)/\partial \lambda(t)$ exist and are continuous, then the optimum solutions will be obtained when the following two equations hold truth, i.e.,
\begin{eqnarray}
\begin{split}
&\frac{\partial \mathcal{L}\big(x(t),\lambda(t),t\big)}{\partial x(t)}=\mathcal{Q}(t)x(t)+\mathcal{P}(t)+\mathcal{J}^{\text{T}}(t)\lambda(t)=0
\\&\frac{\partial \mathcal{L}\big(x(t),\lambda(t),t\big)}{\partial \lambda(t)}=\mathcal{J}(t)x(t)-\mathcal{B}(t)=0.
\label{eqn.Lagrange-2}\
\end{split}
\end{eqnarray}
Equation (\ref{eqn.Lagrange-2}) can be further rewritten into a matrix form as
\begin{eqnarray}
\mathcal{A}(t)\mathcal{Y}(t)=\mathcal{Z}(t)
\label{eqn.matrix-1}
\end{eqnarray}
where
\begin{eqnarray}
\begin{split}
\mathcal{A}(t)&:=
\begin{bmatrix}
\mathcal{Q}(t) &\ \mathcal{J}^{\text{T}}(t)\\
\mathcal{J}(t) &\textbf{0}_{m\times m}
\end{bmatrix}
\in \mathbb{R}^{(n+m)\times(n+m)},
\\ \mathcal{Y}(t)&:=
\begin{bmatrix}
x(t) \\
\lambda(t)
\end{bmatrix}
\in \mathbb{R}^{n+m},
\\ \mathcal{Z}(t)&:=
\begin{bmatrix}
-\mathcal{P}(t) \\
\mathcal{B}(t)
\end{bmatrix}
\in \mathbb{R}^{n+m}.
\label{eqn.simplification}
\end{split}
\end{eqnarray}
$\mathcal{A}(t)$ and $\mathcal{Z}(t)$ are smoothly time-varying coefficient matrix and vector due to the smoothness and continuation of time-varying coefficient matrices $\mathcal{Q}(t)$, $\mathcal{J}(t)$ and vector $\mathcal{B}(t)$; $\mathcal{Y}(t)\in \mathbb{R}^{(n+m)}$ denotes an unknown vector and it needs to be solved at any time instant $t$.

Solving time-varying QP problem (\ref{eqn.QP}) is equivalent to solving the matrix equation (\ref{eqn.matrix-1}). Since the convex QP problem (\ref{eqn.QP}) is time-varying, i.e., coefficient vectors and matrices are changing as time $t$ goes, the theoretical solutions will change all the time. For getting better robustness property, steady states are desired if the time-varying optimal solution is expected to be obtained. In order to obtain better understanding and comparison of the proposed algorithm, the time-varying theoretical solution can be written as
\begin{eqnarray}
\mathcal{Y}^{*}(t)=[x^{*\text{T}}(t),\lambda^{*\text{T}}(t)]^{\text{T}}
=\mathcal{A}^{-1}(t)\mathcal{Z}(t) \in \mathbb{R}^{n+m}.
\label{eqn.matrix-2}
\end{eqnarray}

\textbf{Step 3:}
To obtain the optimum of the matrix equation (\ref{eqn.matrix-1}), a vector-type error function is defined as
\begin{eqnarray}
\epsilon(t)=\mathcal{A}(t)\mathcal{Y}(t)-\mathcal{Z}(t) \in \mathbb{R}^{n+m}.
\label{eqn.error function-1}
\end{eqnarray}

\begin{figure}
  \psfrag{Phi}[c][c][1.19]{$\phi(u)$~~}
  \psfrag{u}[c][c][1.19]{$u$~~~}
  \psfrag{0}[c][c][1.19]{~$0$}
  \includegraphics[width=.99\columnwidth]{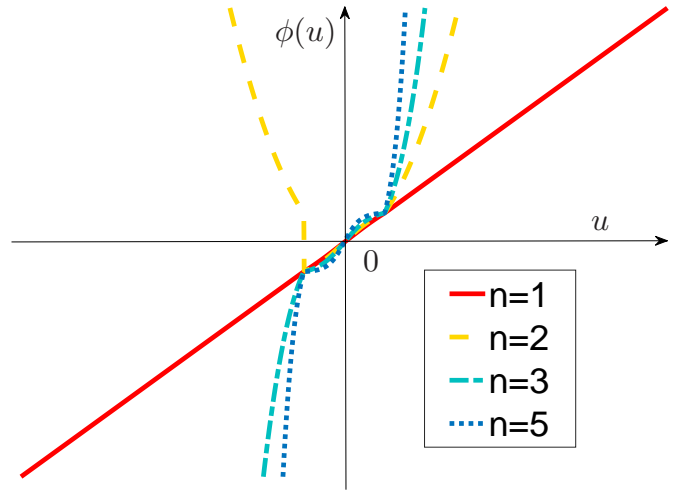}\\
  \caption{The power-sigmoid activation function $\phi(u)$ with different parameter $n$. Evidently such parameter should be odd, and with the increase of $n$, the acceleration speed is enhanced.}\label{fig.AF}
\end{figure}

\textbf{Step 4:}
To make this error function $\epsilon(t)$ approach zero, the negative time derivative of error function $\epsilon(t)$ is necessary. Based on this reason, a neural dynamic design formula is described as
\begin{eqnarray}
\dot\epsilon(t)=\frac{\text{d}\epsilon(t)}{\text{d}t}=-\nu_{1}\Phi(\epsilon(t))-\nu_{2}\int_{0}^{t}\epsilon(\tau)\text{d}\tau
\label{eqn.dot-e}
\end{eqnarray}
where $\nu_{1}>0$ and $\nu_{2}>0$ denote the scalar-valued constant parameters used to scale the convergence rate; $\Phi(\cdot)$ is the activation-function processing-array. Each scalar-valued processing unit $\phi(\cdot)$ of $\Phi(\cdot)$ should be a monotonically-increasing odd activation function. In this paper, a power-sigmoid type activation function (as shown in Fig. \ref{fig.AF}) is leaded to accelerate the convergence, i.e.,\\
$\bullet$ Power-sigmoid activation function:

\begin{eqnarray}
~~~~\phi(u)=\begin{cases}
&\dfrac{1+\exp(-n)}{1-\exp(-n)}\cdot\dfrac{1-\exp(-nu)}{1+\exp(-nu)}~~~\text{if} \ |u|\leq1,\\
&~~~~~~~~~~~~~~~~~u^{n} ~~~~~~~~~~~~~~ \ ~~~~\text{otherwise}.
\end{cases}\nonumber
\end{eqnarray}

It is worth mentioning that the scalar-valued processing unit $\phi(\cdot)$ of $\Phi(\cdot)$ guarantees the following properties:\\
a) If $\epsilon_{i}(t)>0$ or $\epsilon_{i}(t)<0$, then $\epsilon_{i}(t)\phi(\epsilon_{i}(t))>0$;\\
b) If $\epsilon_{i}(t)=0$, then $\epsilon_{i}(t)\phi(\epsilon_{i}(t))=0$.

\textbf{Step 5:}
By expanding the neural dynamic design formula (\ref{eqn.dot-e}), the matrix form of IE-RNN is obtained, i.e.,
\begin{eqnarray}
\begin{split}
\mathcal{A}(t)\mathcal{\dot{Y}}(t)=&-\nu_{1}\Phi(\mathcal{A}(t)\mathcal{Y}(t)-\mathcal{Z}(t))+\mathcal{\dot{Z}}(t)
\\&-\nu_{2}\int_{0}^{t}(\mathcal{A}(t)\mathcal{Y}(t)-\mathcal{Z}(t))\text{d}\tau-\mathcal{\dot{A}}(t)\mathcal{Y}(t).
\end{split}
\label{eqn.IE-RNN}
\end{eqnarray}

According to the traditional zeroing recurrent neural dynamic method \cite{Zhang2010Robustness}, the integration term is omitted. As for QP problem (\ref{eqn.QP}), the matrix form of Z-RNN is
\begin{eqnarray}
\begin{split}
\mathcal{A}(t)\mathcal{\dot{Y}}(t)=&-\nu_{1}\Phi(\mathcal{A}(t)\mathcal{Y}(t)-\mathcal{Z}(t))+\mathcal{\dot{Z}}(t)
\\&-\mathcal{\dot{A}}(t)\mathcal{Y}(t).
\end{split}
\label{eqn.Z-RNN}
\end{eqnarray}

\section{Theoretical Analysis}

\begin{theorem}
(Convergence Theorem)\\
The state vector $\mathcal{Y}(t)$ of IE-RNN (\ref{eqn.IE-RNN}) globally converges to the theoretical solution $\mathcal{Y}^{*}(t)$ (\ref{eqn.matrix-2}) with exponential rate, and the optimal solution to time-varying QP problem (\ref{eqn.QP}) constituted by the first $n$ elements of $\mathcal{Y}(t)$.
\end{theorem}
\begin{IEEEproof}
Firstly, in order to facilitate our following discussion, a scalar type of the neural dynamic design formula (\ref{eqn.dot-e}) is described as
\begin{eqnarray}
\begin{split}
\dot\epsilon_{i}(t)=\frac{\text{d}\epsilon_{i}(t)}{\text{d}t}=-\nu_{1}\phi(\epsilon_{i}(t))
&-\nu_{2}\int_{0}^{t}\epsilon_{i}(\tau)\text{d}\tau, \\& \forall i\in1, ..., n+m.
\label{eqn.scalar-original}
\end{split}
\end{eqnarray}

Based on Lyapunov Theorem \cite{Mazenc2006Further}, by selecting a Lyapunov candidate $\mathcal{V}_{i}(t)=\epsilon_{i}^{2}(t)/2+\nu_{2}(\int_{0}^{t}\epsilon_{i}(\tau)\text{d}\tau)^{2}$, the following result is obtained
\begin{eqnarray}
\mathcal{\dot{V}}_{i}(t)=-\nu_{1}\sum_{i=1}^{n+m}\epsilon_{i}(t)\phi(\epsilon_{i}(t)).
\end{eqnarray}

With $\mathcal{V}_{i}(t)>0$ and $\mathcal{\dot{V}}_{i}(t)<0$, we can draw the conclusion that the $i$-th system (\ref{eqn.scalar-original}) globally converges to zero. In other words, the state vector $\mathcal{Y}(t)$ of IE-RNN (\ref{eqn.IE-RNN}) globally converges to the theoretical solution $\mathcal{Y}^{*}(t)$, of which the optimal solution constituted by the first $n$ elements to time-varying QP problem (\ref{eqn.QP}).

Secondly, considering $\vartheta(t)=\int_{0}^{t}\epsilon(\tau)\text{d}\tau$, let $\vartheta_{i}(t)$, $\dot{\vartheta}_{i}(t)$, $\ddot{\vartheta}_{i}(t)$, $\epsilon_{i}(t)$ be the $i$-th elements of $\vartheta(t)$, $\dot{\vartheta}(t)$, $\ddot{\vartheta}(t)$, $\epsilon(t)$, respectively. The scalar-valued design formula (\ref{eqn.scalar-original}) is reformulated as $\ddot{\vartheta}_{i}(t)=-\nu_{1}\dot{\vartheta}_{i}(t)-\nu_{2}\vartheta_{i}(t)$, and the solutions are
\begin{eqnarray}
\begin{cases}
\delta_{1}=\big(-\nu_{1}+\sqrt{\nu_{1}^{2}-4\nu_{2}}~\big)/2\\
\delta_{2}=\big(-\nu_{1}-\sqrt{\nu_{1}^{2}-4\nu_{2}}~\big)/2
\end{cases}.
\end{eqnarray}

Assuming that initial values $\vartheta_{i}(0)$ and $\dot{\vartheta}_{i}(0)$ satisfy $\vartheta_{i}(0)=0$ and $\dot{\vartheta}_{i}(0)=\epsilon_{i}(0)$, there exists three cases.

\textbf{Case 1:} If $\nu_{1}^{2}>4\nu_{2}$, with $\delta_{1}\neq\delta_{2}$, we have
\begin{eqnarray}
\epsilon_i(t)=\frac{\epsilon_{i}(0)\delta_{1}\exp(\delta_{1}t)
-\epsilon_{i}(0)\delta_{2}\exp(\delta_{2}t)}{\sqrt{\nu_{1}^{2}-4\nu_{2}}}.
\end{eqnarray}

\textbf{Case 2:} If $\nu_{1}^{2}=4\nu_{2}$, with $\delta_{1}=\delta_{2}=-\nu_{1}/2$, we have
\begin{eqnarray}
\epsilon_i(t)=(1+\delta_{1}t)\epsilon_{i}(0)\exp(\delta_{1}t).
\end{eqnarray}

\textbf{Case 3:} If $\nu_{1}^{2}<4\nu_{2}$, with conjugate complexes $\nu_{1}=\alpha+i\beta$ and $\nu_{2}=\alpha-i\beta$, we have
\begin{eqnarray}
\epsilon_i(t)=\epsilon_{i}(0)\exp(\alpha t)\big(\frac{\alpha}{\beta}\sin(\beta t)+\cos(\beta t)\big).
\end{eqnarray}

Hence it leads to the conclusion that starting from randomly initial state $\epsilon(0)$, the residual error $\epsilon(t)$ of IE-RNN (\ref{eqn.IE-RNN}) globally and exponentially converges to zero. That is to say, the state vector $\mathcal{Y}(t)$ of IE-RNN (\ref{eqn.IE-RNN}) globally converges to the theoretical solution $\mathcal{Y}^{*}(t)$ (\ref{eqn.matrix-2}) with exponential rate. The proof is thus complected.
\end{IEEEproof}

\begin{remark}
For time-varying QP problem (\ref{eqn.QP}), the state vector $\mathcal{Y}(t)$ of Z-RNN (\ref{eqn.Z-RNN}), starting from any initial state $\mathcal{Y}(0)$, could globally converge to the unique solution $\mathcal{Y}^*(t)$ of time-varying system (\ref{eqn.matrix-1}). In addition, the first $n$ elements of solution $\mathcal{Y}^*(t)$ constitute the time-varying optimal solution $x^*(t)$ to time-varying QP problem (\ref{eqn.QP}).
\end{remark}

\begin{IEEEproof}
See Theorem 1 in Ref. \cite{Zhang2009Zhang}.
\end{IEEEproof}

In actual hardware implementation, measurement noise often exists, thus the robustness of a noise-polluted system is worth considering. The noise-polluted model of IE-RNN (\ref{eqn.IE-RNN}) is described as
\begin{eqnarray}
\begin{split}
&\mathcal{A}(t)\mathcal{\dot{Y}}(t)=-\nu_{1}\Phi(\mathcal{A}(t)\mathcal{Y}(t)-\mathcal{Z}(t))+\mathcal{\dot{Z}}(t)
\\&-\nu_{2}\int_{0}^{t}(\mathcal{A}(t)\mathcal{Y}(t)-\mathcal{Z}(t))\text{d}\tau-\mathcal{\dot{A}}(t)\mathcal{Y}(t)+\Delta\mathcal{N}
\end{split}
\label{eqn.perturbed}
\end{eqnarray}
where $\Delta\mathcal{N}\in\mathbb{R}^{n+m}$ denotes measurement noise. It is worth mentioning that noise $\Delta\mathcal{N}$ contains the following forms\\
$\bullet$ Constant noise $\Delta\mathcal{N}_{1}(t)=\Delta\mathcal{N}\in\mathbb{R}^{n+m}$.\\
$\bullet$ Time-varying noise $\Delta\mathcal{N}_{2}(t)=t\cdot\Delta\mathcal{N}\in\mathbb{R}^{n+m}$.

\begin{theorem}
(Robustness Theorem)\\
Consider noise-polluted IE-RNN (\ref{eqn.perturbed}), Case 1: with constant noise $\Delta\mathcal{N}_{1}(t)$, the state vector $\mathcal{Y}(t)$ of IE-RNN (\ref{eqn.IE-RNN}) globally converges to the theoretical solution $\mathcal{Y}^{*}(t)$ (\ref{eqn.matrix-2}). Case 2: with time-varying noise $t\cdot\Delta\mathcal{N}$, the state vector $\mathcal{Y}(t)$ of IE-RNN (\ref{eqn.IE-RNN}) converges to the theoretical solution $\mathcal{Y}^{*}(t)$ (\ref{eqn.matrix-2}) with steady residual error $\|\Delta\mathcal{N}/\nu_2\|$.
\end{theorem}

\begin{figure*}
  \centering
  \psfrag{Expected Path}[c][c][0.45]{\textmd{Expected Path}}
  \psfrag{Actual Trajectories}[c][c][0.45]{\textmd{Actual Trajectories}}
  \psfrag{joint-velocity (rad/s)}[c][c][0.45]{~~~~~~\textmd{Joint-velocity (rad/s)}}
  \psfrag{joint-acceleration (rad/s2)}[c][c][0.45]{~~~~~~\textmd{Joint-acceleration (rad/s$^2$)}}
  \psfrag{t (s)}[c][c][0.45]{$t$ \text{(s)}~}
  \psfrag{X (m)}[c][c][0.45]{$X$ \text{(m)}}
  \psfrag{Y (m)}[c][c][0.45]{~~$Y$ \text{(m)}}
  \psfrag{eX(t)}[c][c][0.43]{$\epsilon_{X}(t)$}
  \psfrag{eY(t)}[c][c][0.43]{$\epsilon_{Y}(t)$}
  \psfrag{eZ(t)}[c][c][0.43]{$\epsilon_{Z}(t)$}
  \psfrag{e(t) (m)}[c][c][0.45]{~~$\epsilon(t)$ \text{(m)}}
  \psfrag{||e(t)||2 (m)}[c][c][0.45]{~~$||\epsilon(t)||_{2}$ \text{m}}
  \psfrag{dq1}[c][c][0.45]{~$\dot{\theta}_1$}
  \psfrag{dq2}[c][c][0.45]{~$\dot{\theta}_2$}
  \psfrag{dq3}[c][c][0.45]{~$\dot{\theta}_3$}
  \psfrag{dq4}[c][c][0.45]{~$\dot{\theta}_4$}
  \psfrag{dq5}[c][c][0.45]{~$\dot{\theta}_5$}
  \psfrag{dq6}[c][c][0.45]{~$\dot{\theta}_6$}
  \psfrag{ddq1}[c][c][0.45]{~$\ddot{\theta}_1$}
  \psfrag{ddq2}[c][c][0.45]{~$\ddot{\theta}_2$}
  \psfrag{ddq3}[c][c][0.45]{~$\ddot{\theta}_3$}
  \psfrag{ddq4}[c][c][0.45]{~$\ddot{\theta}_4$}
  \psfrag{ddq5}[c][c][0.45]{~$\ddot{\theta}_5$}
  \psfrag{ddq6}[c][c][0.45]{~$\ddot{\theta}_6$}
  \subfigure []
  {\includegraphics[width=0.195\textwidth]{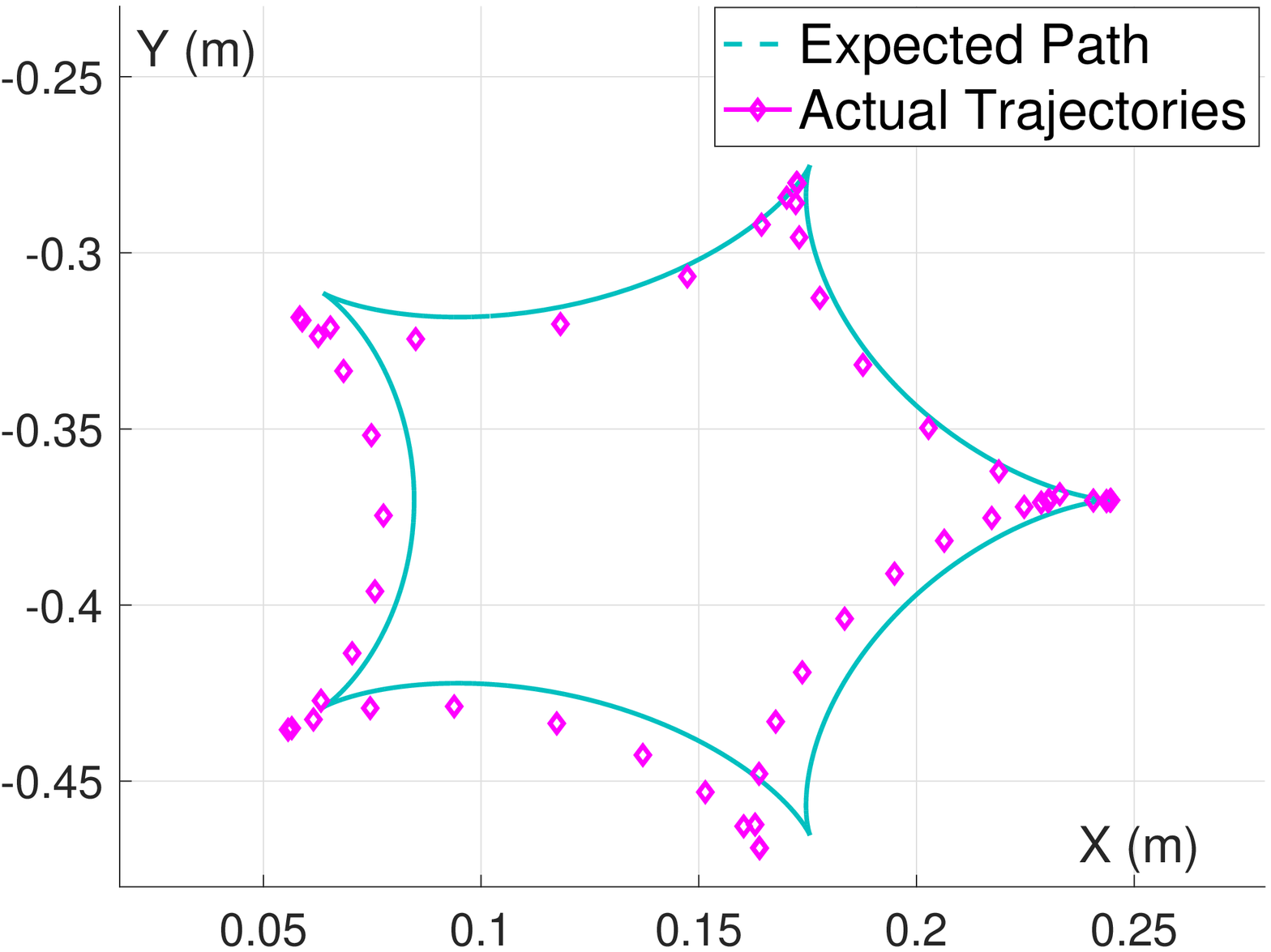}\label{fig.RM-1}}
  \subfigure []
  {\includegraphics[width=0.195\textwidth]{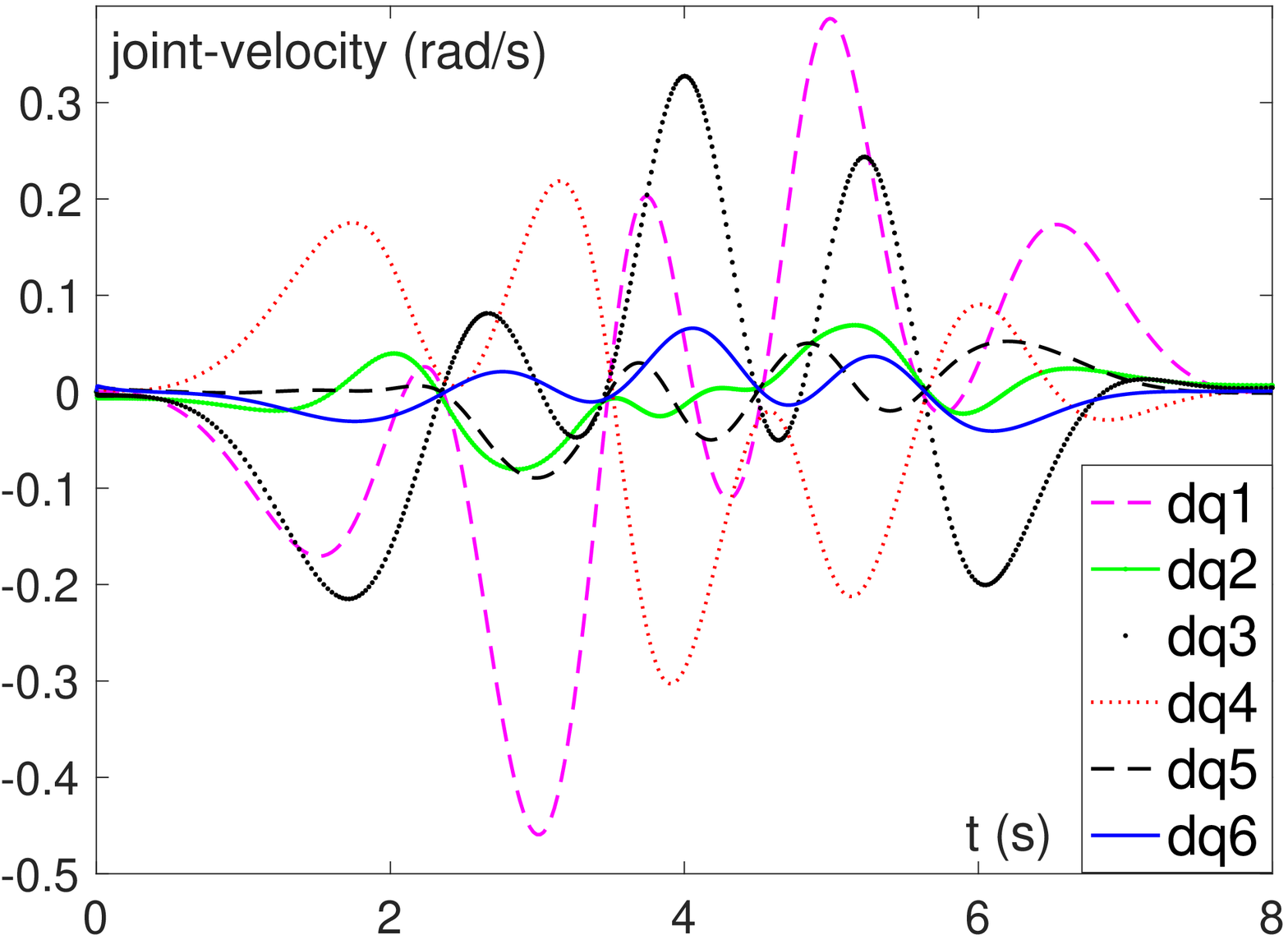}\label{fig.RM-2}}
  \subfigure []
  {\includegraphics[width=0.195\textwidth]{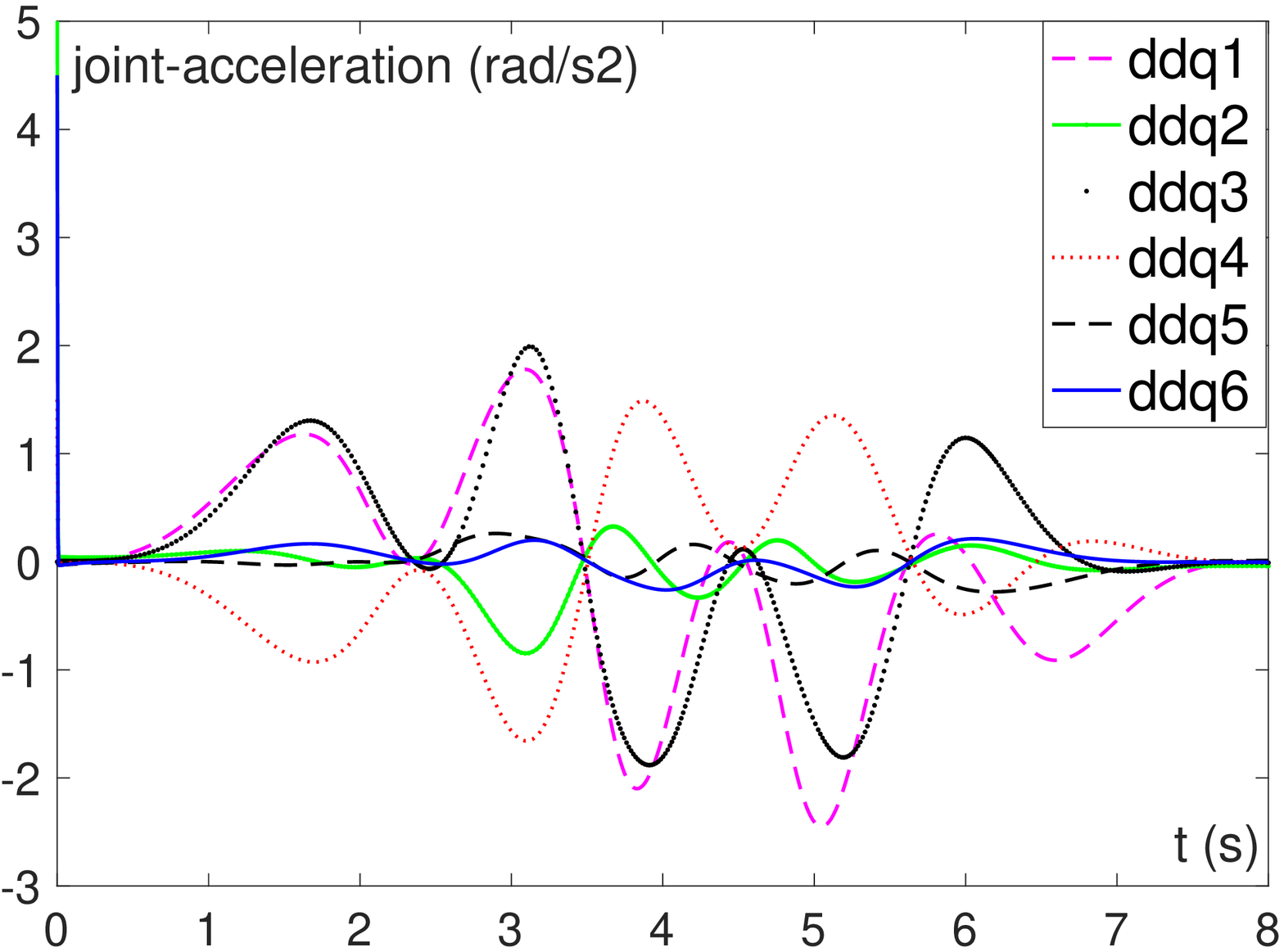}\label{fig.RM-3}}
  \subfigure []
  {\includegraphics[width=0.195\textwidth]{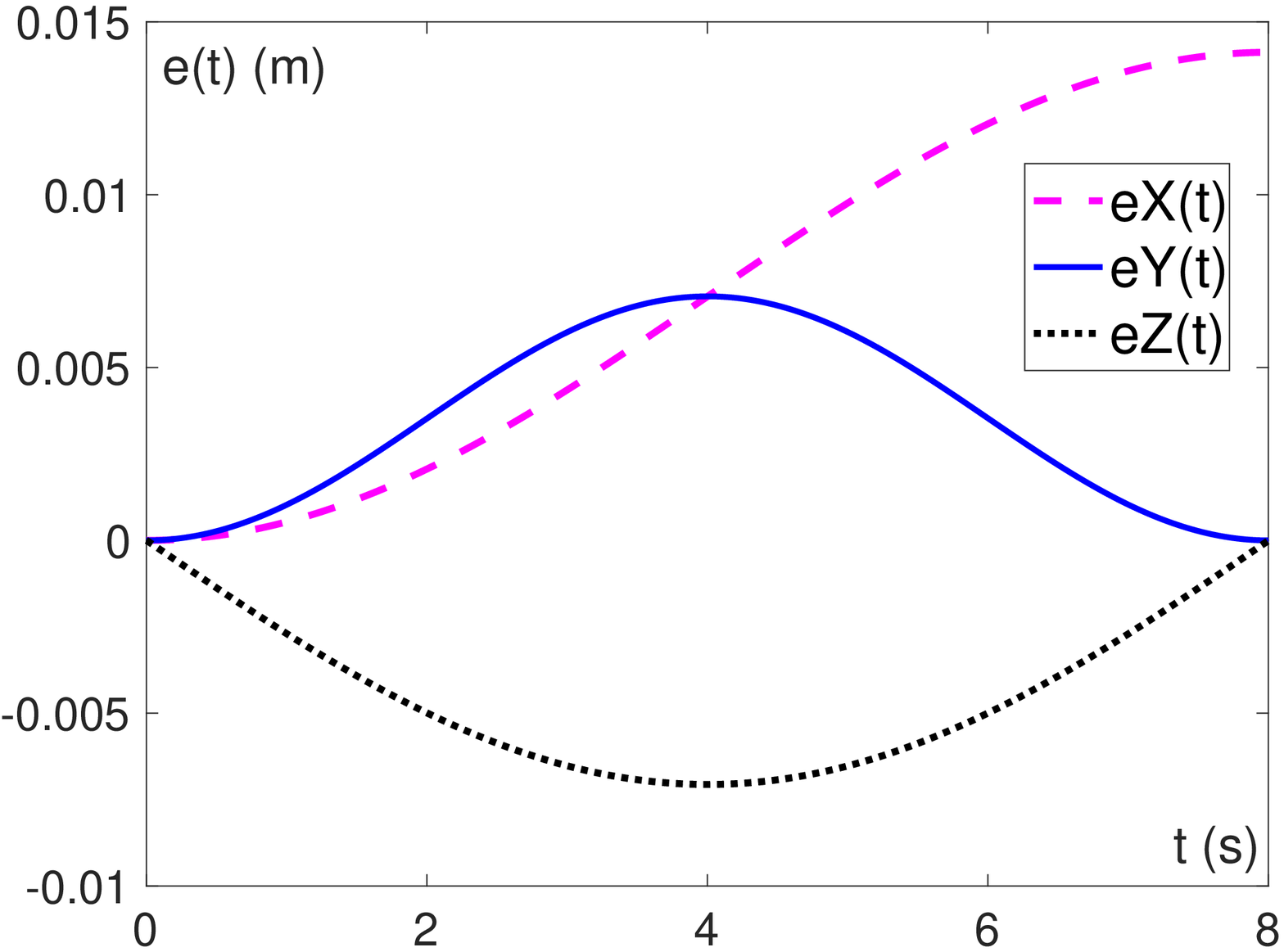}\label{fig.RM-4}}
  \subfigure []
  {\includegraphics[width=0.195\textwidth]{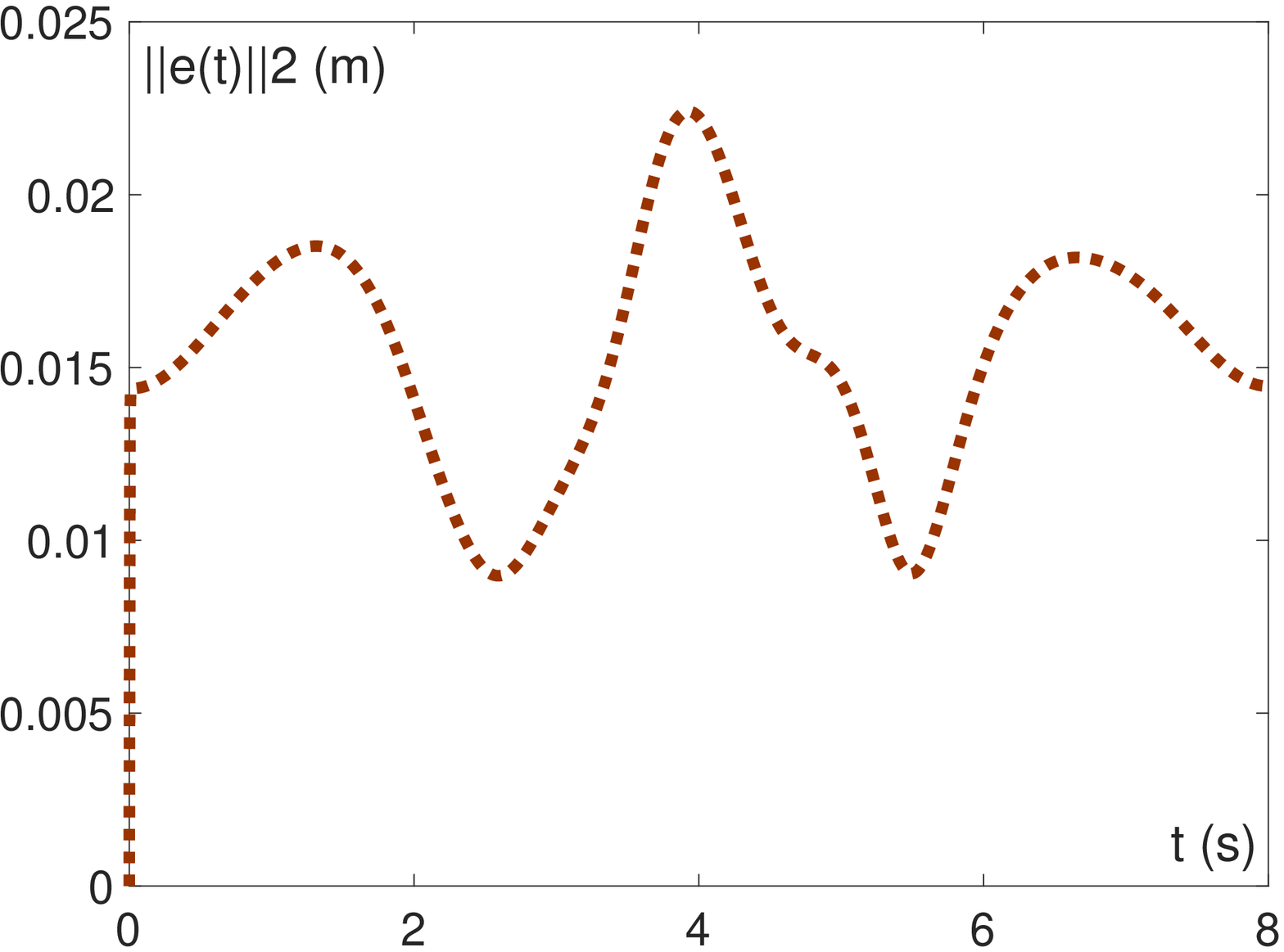}\label{fig.RM-5}}\\
  \subfigure []
  {\includegraphics[width=0.195\textwidth]{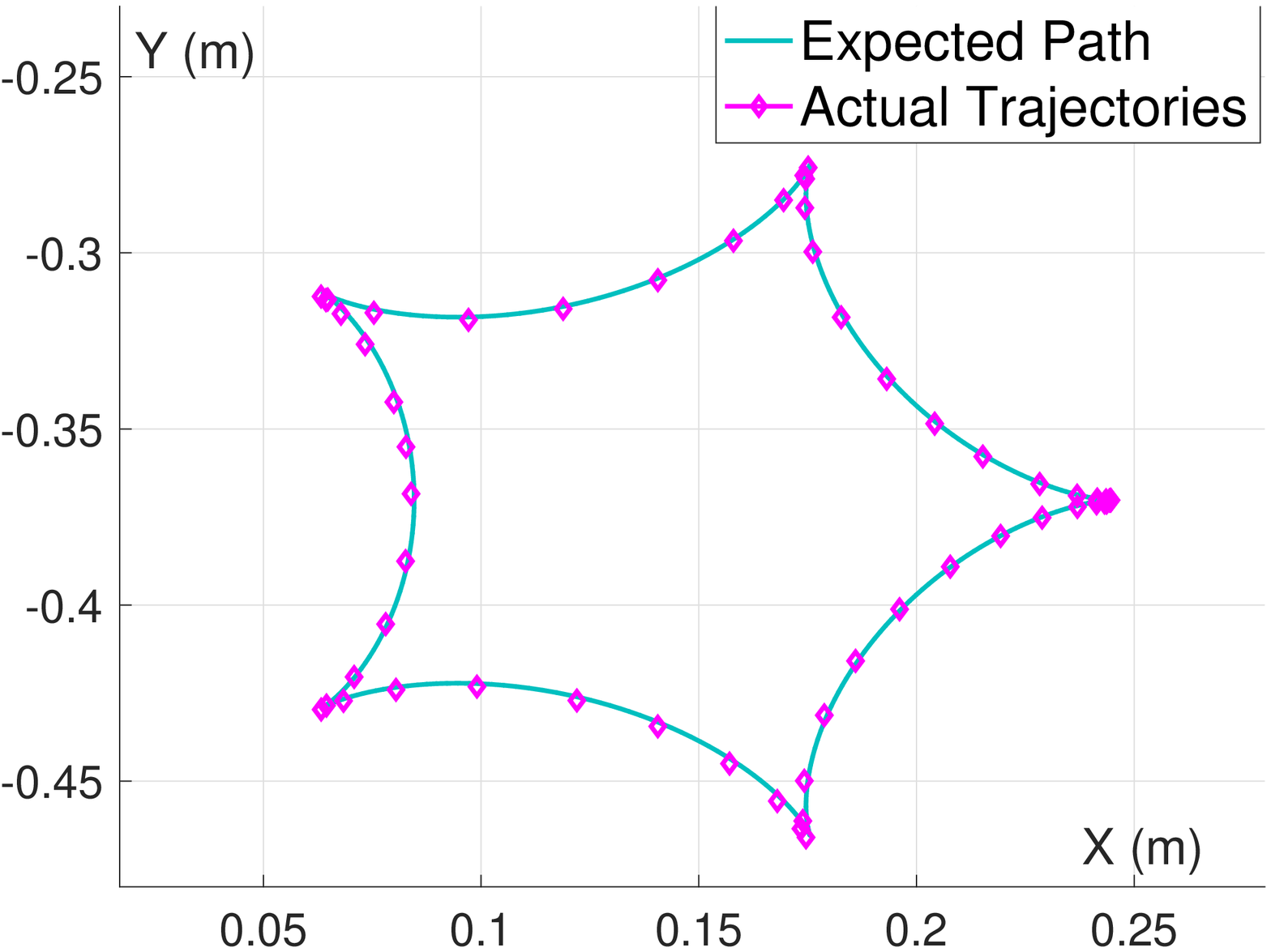}\label{fig.RM-6}}
  \subfigure []
  {\includegraphics[width=0.195\textwidth]{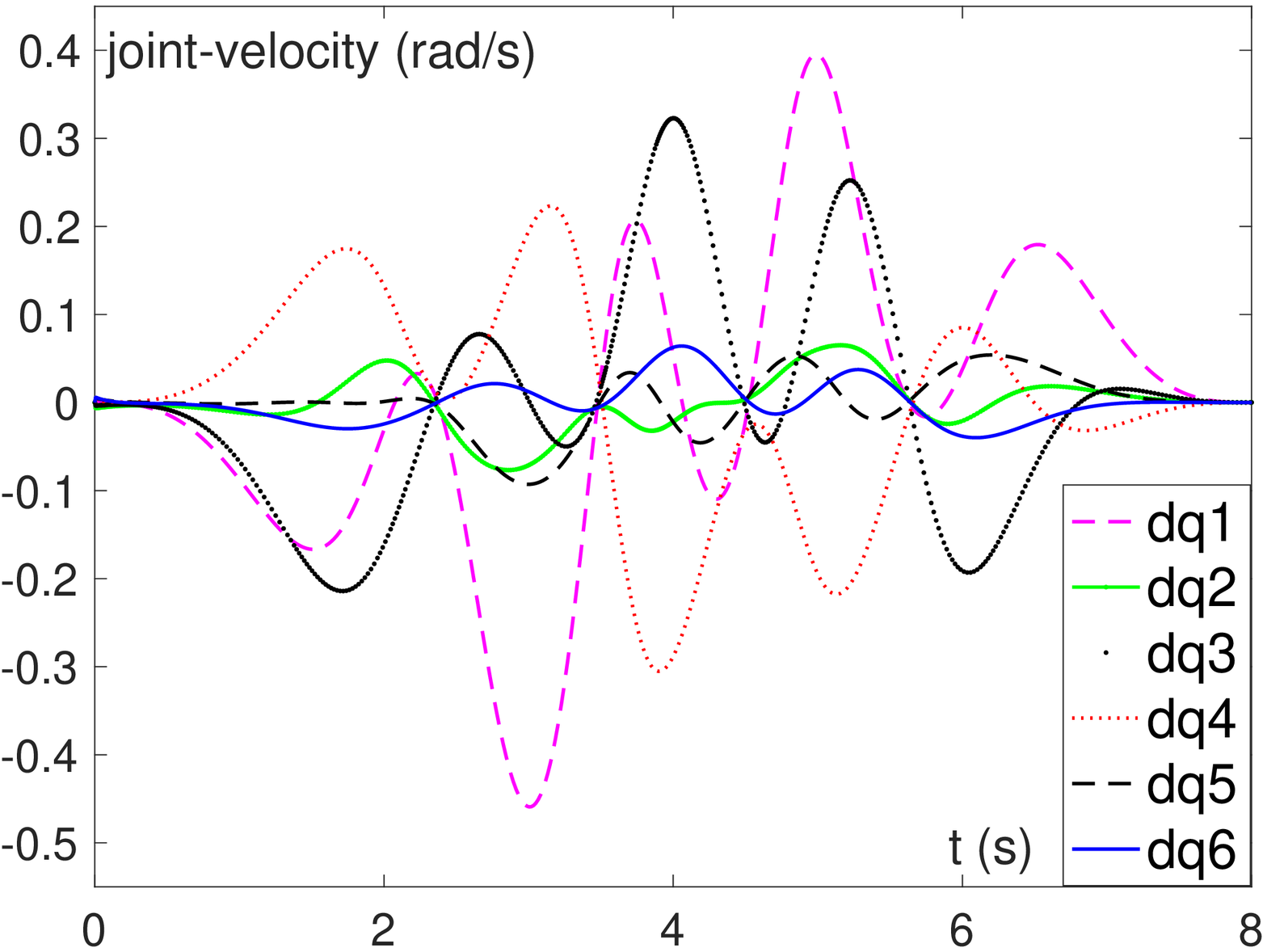}\label{fig.RM-7}}
  \subfigure []
  {\includegraphics[width=0.195\textwidth]{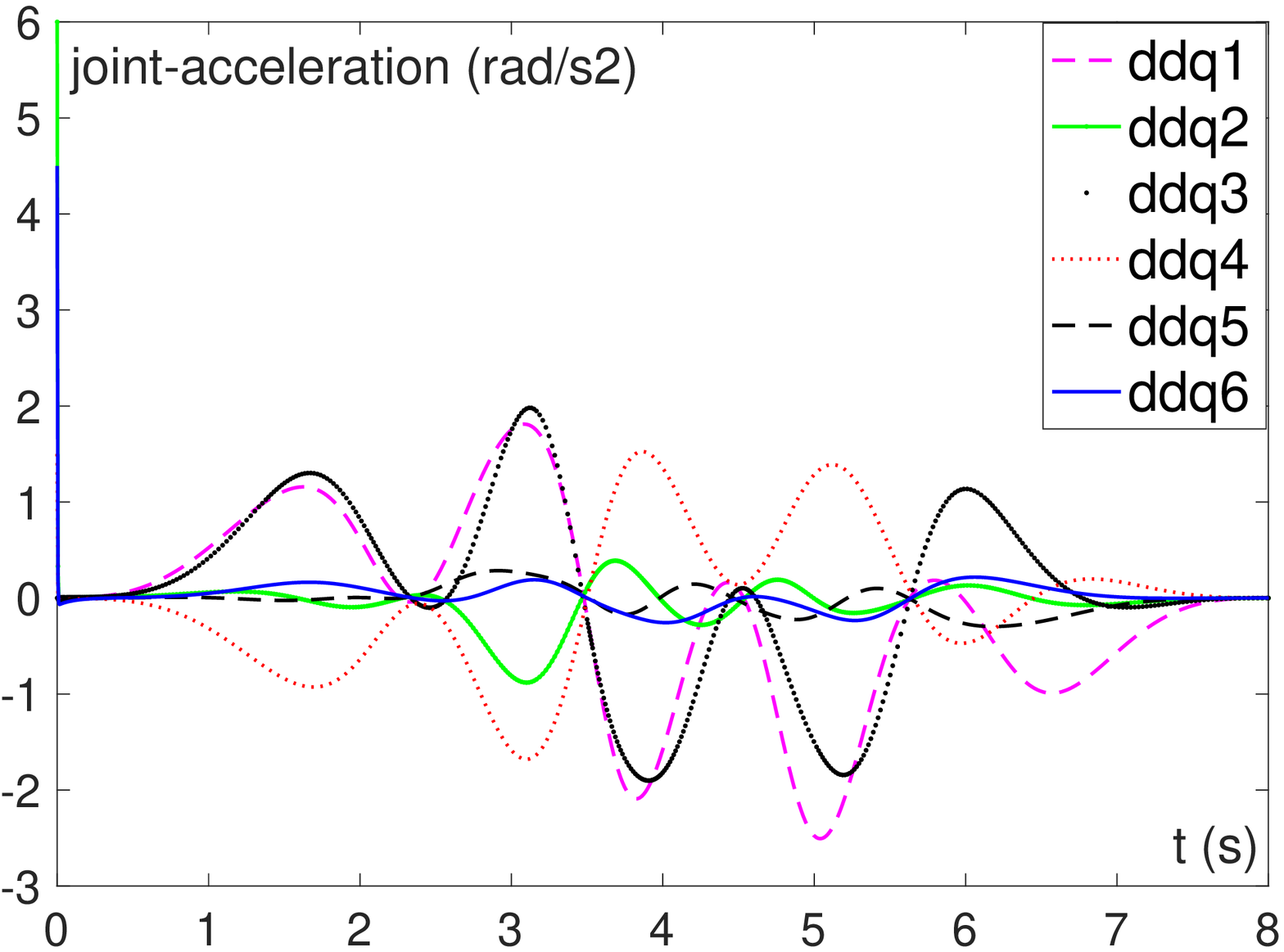}\label{fig.RM-8}}
  \subfigure []
  {\includegraphics[width=0.195\textwidth]{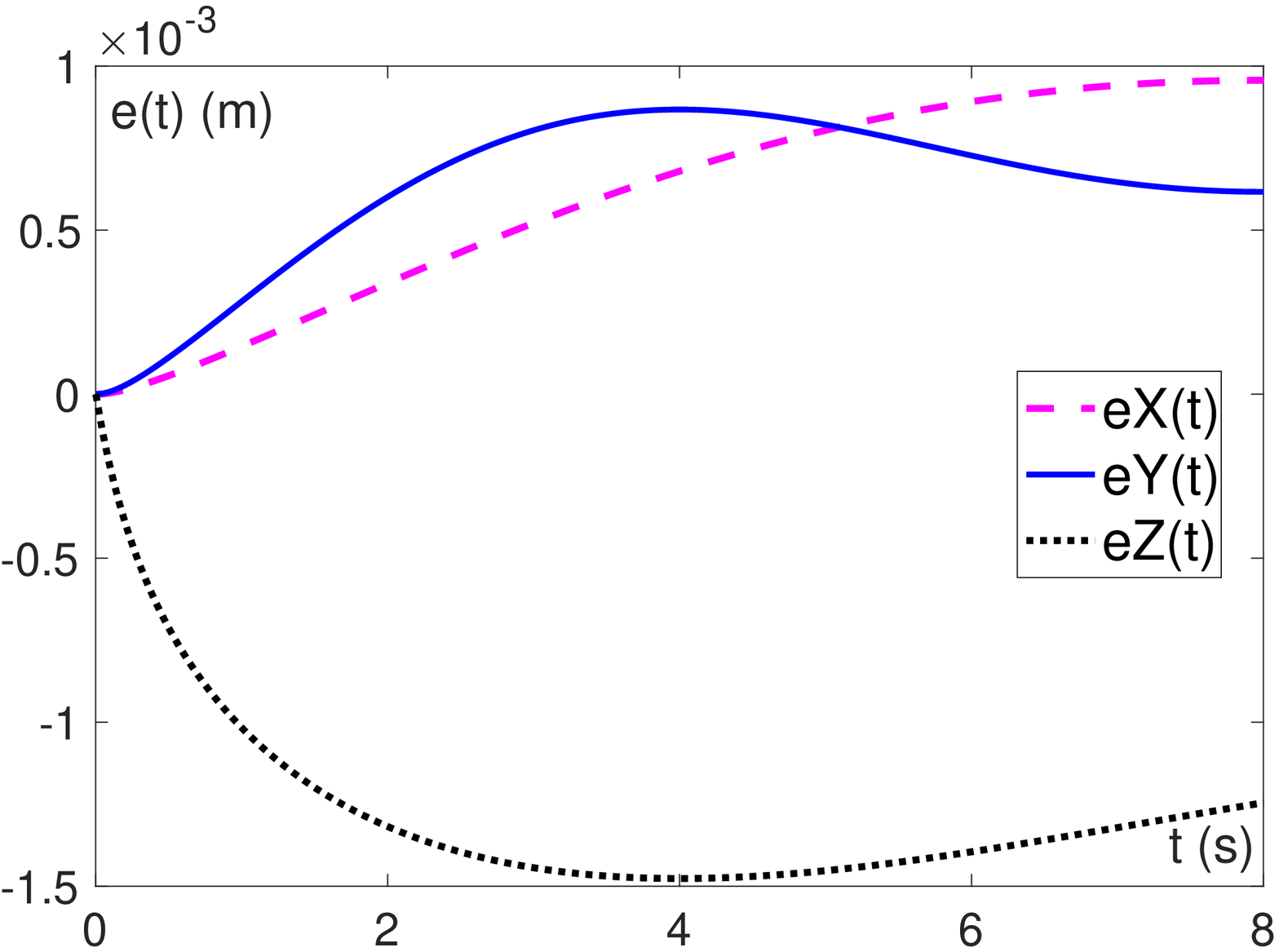}\label{fig.RM-9}}
  \subfigure []
  {\includegraphics[width=0.195\textwidth]{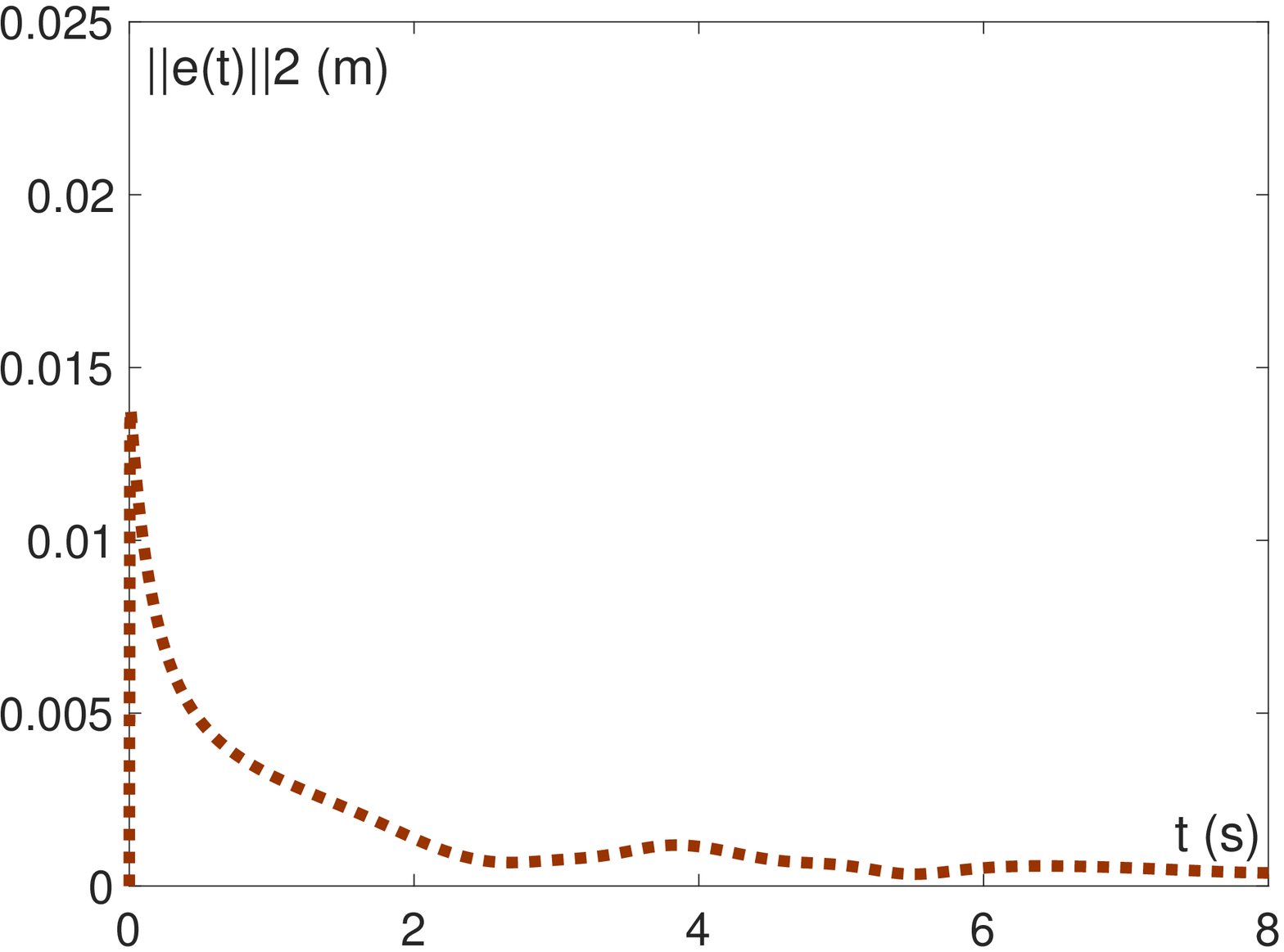}\label{fig.RM-10}}
  \caption{Computer simulation results via Repetitive Motion Scheme when tracking a starfish path. (a) Trajectories of Z-RNN. (b) Joint-velocity of Z-RNN. (c) Joint-acceleration of Z-RNN. (d) Position error of Z-RNN. (e) RMS error of IE-RNN. (f) Trajectories of IE-RNN. (g) Joint-velocity of IE-RNN. (h) Joint-acceleration of IE-RNN. (i) Position error of IE-RNN. (j) RMS error of IE-RNN.}
  \label{fig.EXP1}
\end{figure*}

\begin{IEEEproof}
For the convenience of our discussion and make it easier to understand, the parameter $n$ of power-sigmoid activation function $\Phi(\cdot)$ is set as $n=1$, and other conditions are analogous. Firstly, considering constant noise $\Delta\mathcal{N}_{1}(t)$, by using Laplace transform, the scalar type of noise-polluted IE-RNN is reformulated as
\begin{eqnarray}
s\epsilon_{i}(s)=\epsilon_{i}(0)-\nu_{1}\epsilon_{i}(s)-\frac{\nu_{2}}{s}\epsilon_{i}(s)+\Delta\mathcal{N}_{i}(s)
\end{eqnarray}
which can be further simplified as
\begin{eqnarray}
\epsilon_{i}(s)=\frac{s\epsilon_{i}(0)+s\Delta\mathcal{N}_{i}(s)}{s^2+s\nu_1+\nu_2}.
\label{eqn.s-system}
\end{eqnarray}

The transfer function of system (\ref{eqn.s-system}) is $s/(s^2+s\nu_1+\nu_2)$, and the poles are $s_1=(-\nu_1+\sqrt{\nu_1^2-4\nu_2})/2$ and $s_2=(-\nu_1-\sqrt{\nu_1^2-4\nu_2})/2$. Based on Nyquist's Theorem \cite{Luse1988A}, with positive numbers $\nu_1$ and $\nu_2$, poles $s_1$ and $s_2$ locate on the left half plane, which implies that system (\ref{eqn.s-system}) is stable. By introducing Final-value Theorem \cite{Luse1988A}, the following conclusions are obtained, i.e.,

\textbf{Case 1:} For constant noise $\Delta\mathcal{N}_{1}(t)$, $\mathcal{N}_{i}(s)=\Delta\mathcal{N}_{i}(t)/s$, hence
\begin{eqnarray}
\lim_{t\rightarrow\infty}\epsilon_{i}(t)=\lim_{s\rightarrow0}s\epsilon_{i}(s)
=\lim_{s\rightarrow0}\frac{s^2\epsilon_{i}(0)+s\Delta\mathcal{N}_i}{s^2+s\nu_1+\nu_2}=0.
\end{eqnarray}

With $\lim_{s\rightarrow0}\|\epsilon(t)\|_2=0$, we can draw the conclusion that state vector $\mathcal{Y}(t)$ of IE-RNN (\ref{eqn.IE-RNN}) globally converges to the theoretical solution $\mathcal{Y}^{*}(t)$ (\ref{eqn.matrix-2}).

\textbf{Case 2:} For time-varying noise $t\cdot\Delta\mathcal{N}_{1}(t)$, $\mathcal{N}_{i}(s)=\Delta\mathcal{N}_{i}(t)/s^2$, hence
\begin{eqnarray}
\lim_{t\rightarrow\infty}\epsilon_{i}(t)=\lim_{s\rightarrow0}s\epsilon_{i}(s)
=\lim_{s\rightarrow0}\frac{s^2\epsilon_{i}(0)+\Delta\mathcal{N}_i}{s^2+s\nu_1+\nu_2}
=\frac{\Delta\mathcal{N}_i}{\nu_2}.
\end{eqnarray}

With $\lim_{s\rightarrow0}\|\epsilon(t)\|_2=\|\Delta\mathcal{N}_i\|_2/\nu_2$, we can draw the conclusion that the residual error is upper-bounded and converges to zero when $\nu_2$ tends to $\infty$. The proof is thus complected.

\end{IEEEproof}

\section{Experiments}
In this section, we make computer simulations and practical experiments to verify the authenticity and validity of our aforementioned theoretical analysis and discussion.

The computer simulations are performed with MATLAB R2017b, on a MacBook Pro (2017) with Intel Core i7 CPU at 2.8GHz, 2133MHz LPDDR3 and 16GB of RAM. The practical experiments are performed with a Kinova JACO$^{2}$ robot manipulator, which contains six degrees-of-freedom (DOF).

\begin{figure*}
  \centering
  \psfrag{Expected Path}[c][c][0.45]{\textmd{Expected Path}}
  \psfrag{Actual Trajectories}[c][c][0.45]{\textmd{Actual Trajectories}}
  \psfrag{joint-velocity (rad/s)}[c][c][0.45]{~~~~~~\textmd{Joint-velocity (rad/s)}}
  \psfrag{joint-acceleration (rad/s2)}[c][c][0.45]{~~~~~~\textmd{Joint-acceleration (rad/s$^2$)}}
  \psfrag{t (s)}[c][c][0.45]{$t$ \text{(s)}~}
  \psfrag{X (m)}[c][c][0.45]{$X$ \text{(m)}}
  \psfrag{Y (m)}[c][c][0.45]{~~$Y$ \text{(m)}}
  \psfrag{eX(t)}[c][c][0.43]{$\epsilon_{X}(t)$}
  \psfrag{eY(t)}[c][c][0.43]{$\epsilon_{Y}(t)$}
  \psfrag{eZ(t)}[c][c][0.43]{$\epsilon_{Z}(t)$}
  \psfrag{e(t) (m)}[c][c][0.45]{~~$\epsilon(t)$ \text{(m)}}
  \psfrag{||e(t)||2 (m)}[c][c][0.45]{~~$||\epsilon(t)||_{2}$ \text{m}}
  \psfrag{dq1}[c][c][0.45]{~$\dot{\theta}_1$}
  \psfrag{dq2}[c][c][0.45]{~$\dot{\theta}_2$}
  \psfrag{dq3}[c][c][0.45]{~$\dot{\theta}_3$}
  \psfrag{dq4}[c][c][0.45]{~$\dot{\theta}_4$}
  \psfrag{dq5}[c][c][0.45]{~$\dot{\theta}_5$}
  \psfrag{dq6}[c][c][0.45]{~$\dot{\theta}_6$}
  \psfrag{ddq1}[c][c][0.45]{~$\ddot{\theta}_1$}
  \psfrag{ddq2}[c][c][0.45]{~$\ddot{\theta}_2$}
  \psfrag{ddq3}[c][c][0.45]{~$\ddot{\theta}_3$}
  \psfrag{ddq4}[c][c][0.45]{~$\ddot{\theta}_4$}
  \psfrag{ddq5}[c][c][0.45]{~$\ddot{\theta}_5$}
  \psfrag{ddq6}[c][c][0.45]{~$\ddot{\theta}_6$}
  \subfigure []
  {\includegraphics[width=0.195\textwidth]{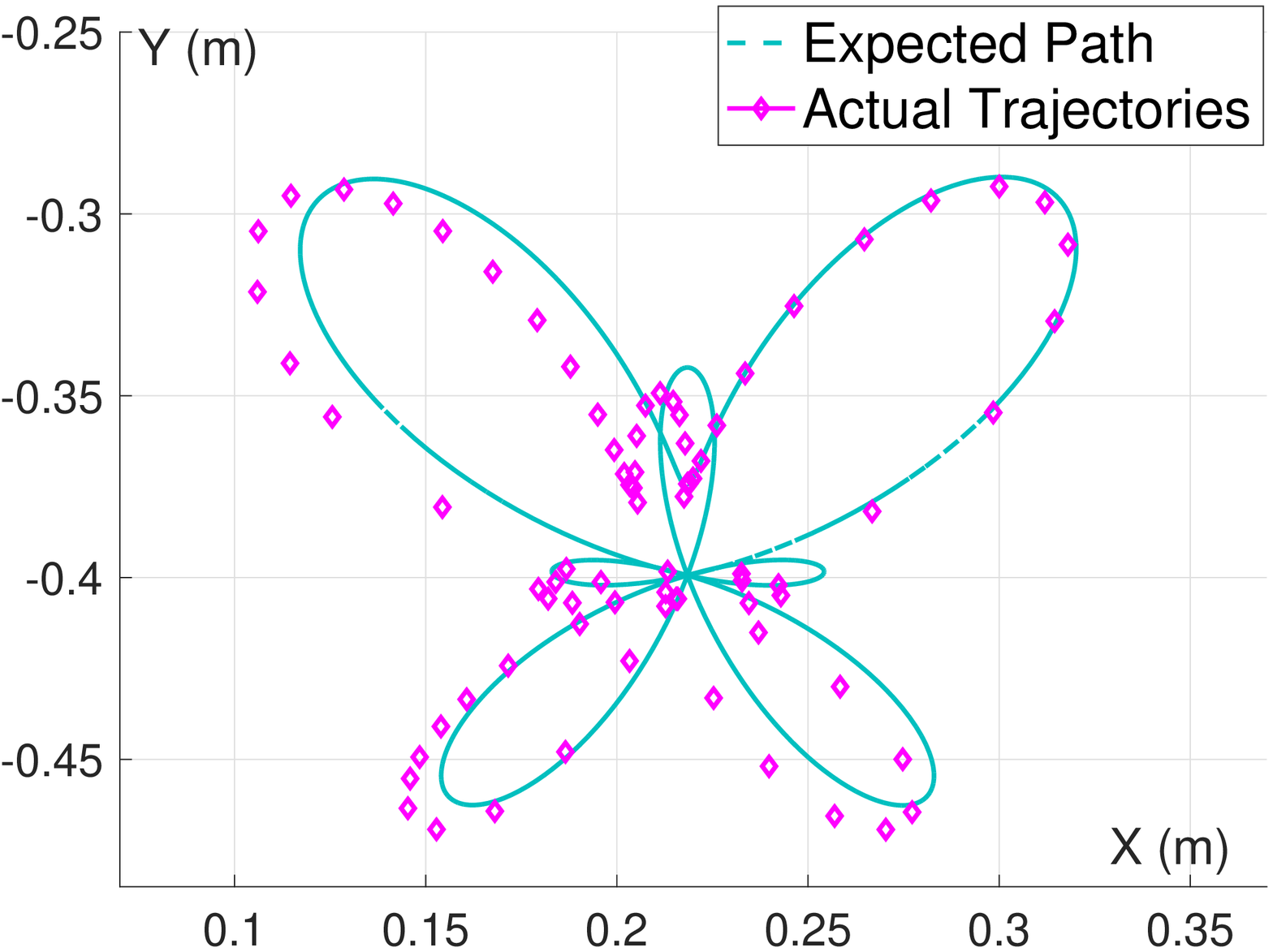}\label{fig.RM-1}}
  \subfigure []
  {\includegraphics[width=0.195\textwidth]{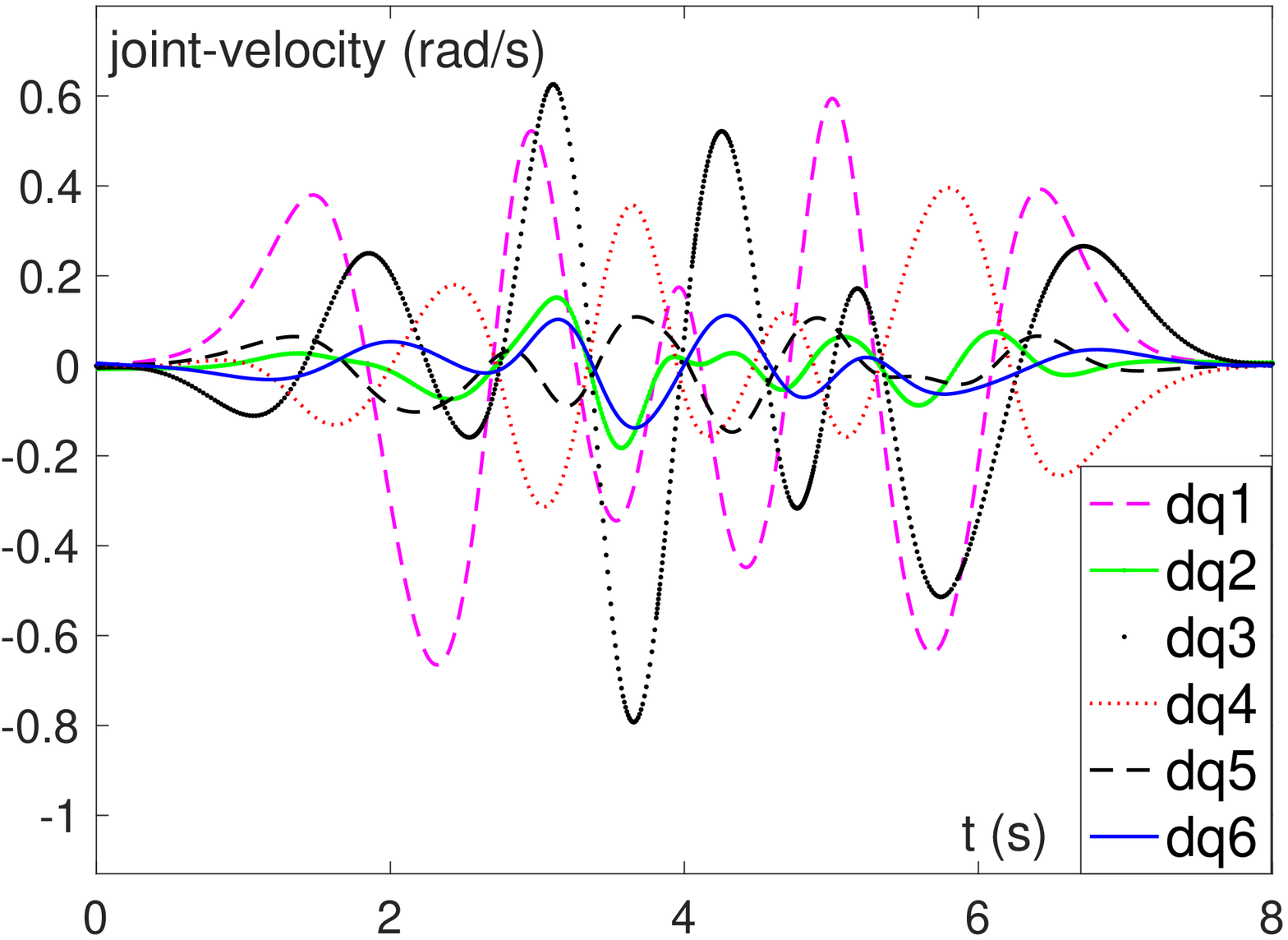}\label{fig.RM-2}}
  \subfigure []
  {\includegraphics[width=0.195\textwidth]{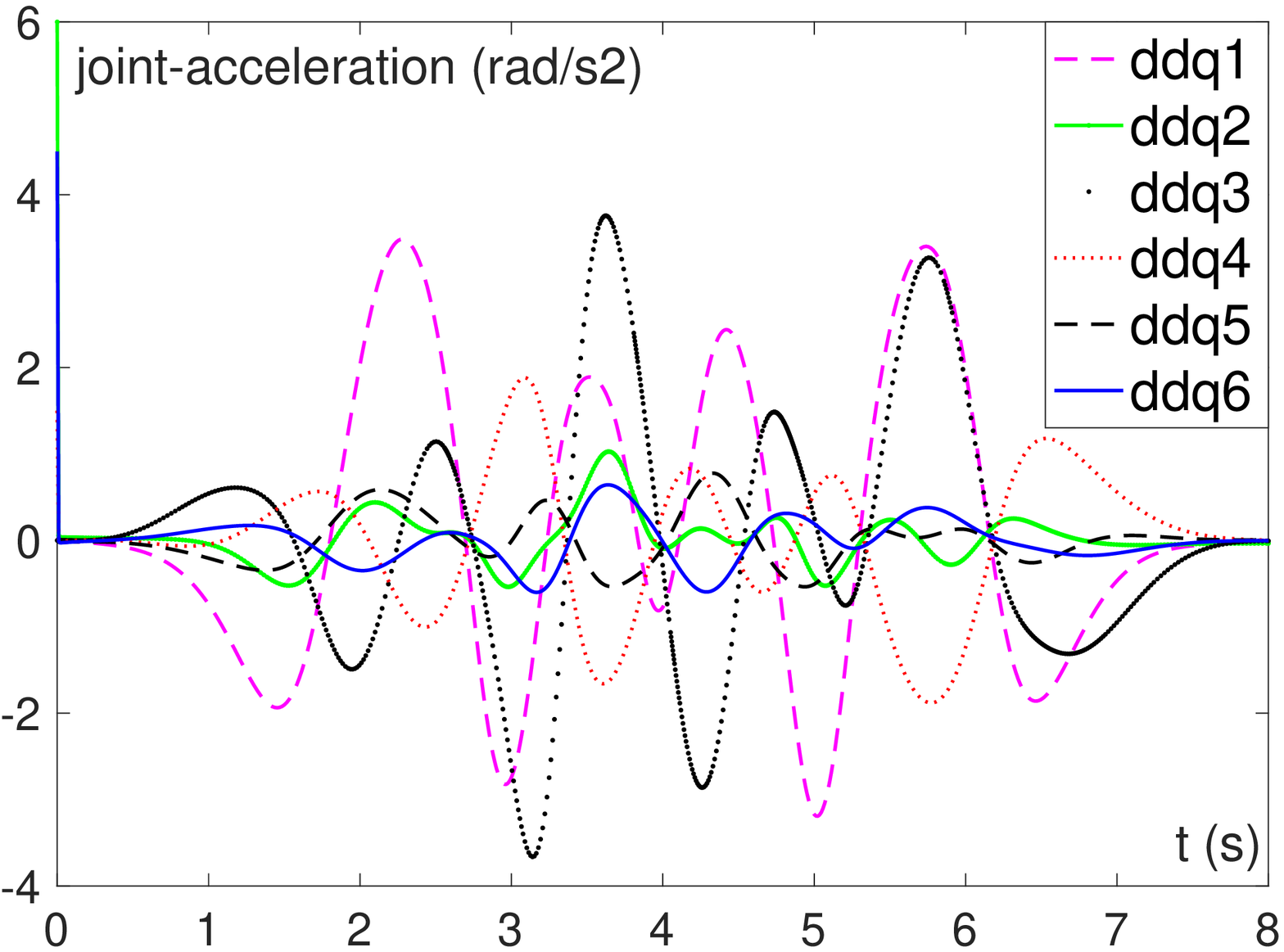}\label{fig.RM-3}}
  \subfigure []
  {\includegraphics[width=0.195\textwidth]{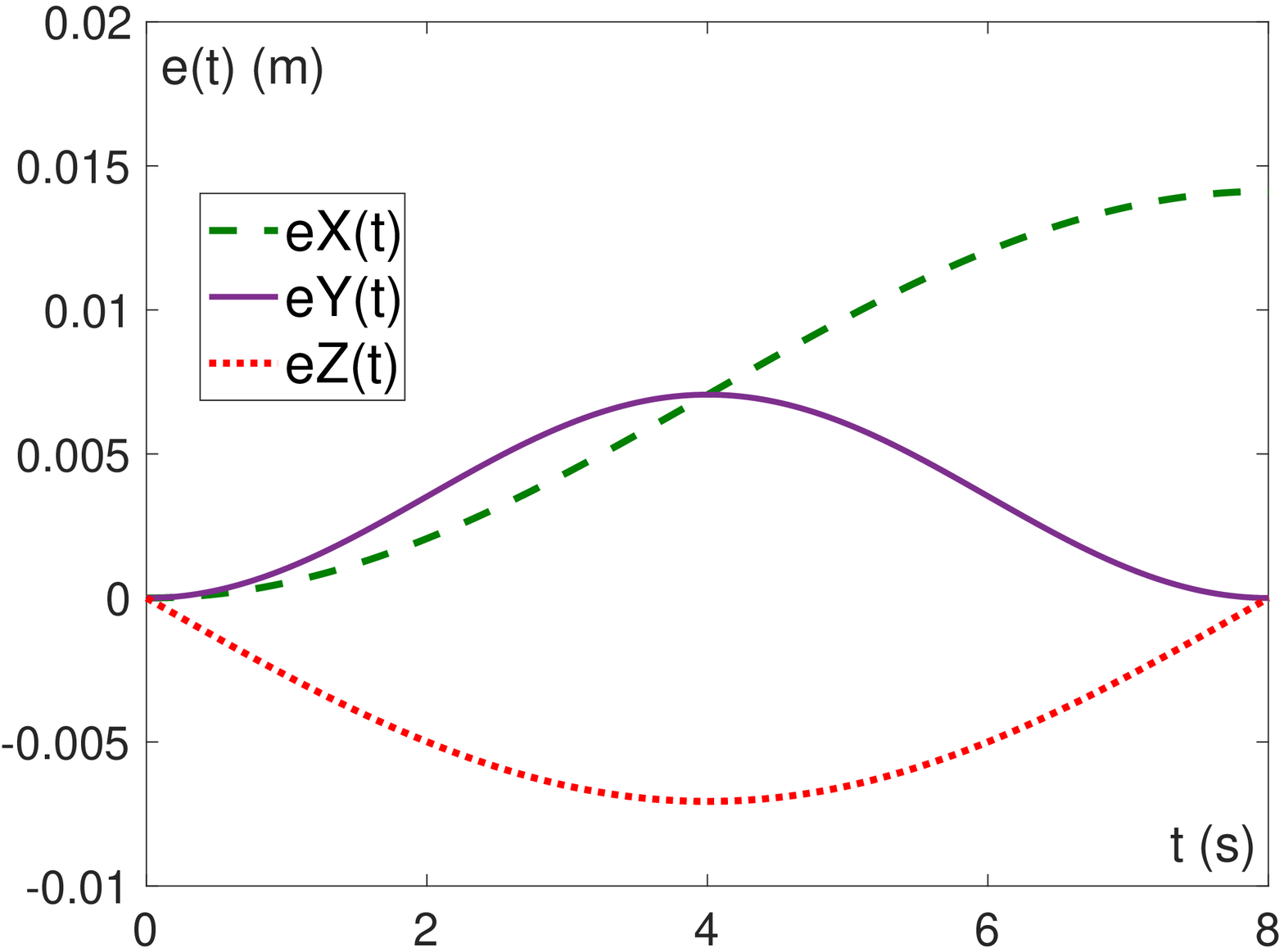}\label{fig.RM-4}}
  \subfigure []
  {\includegraphics[width=0.195\textwidth]{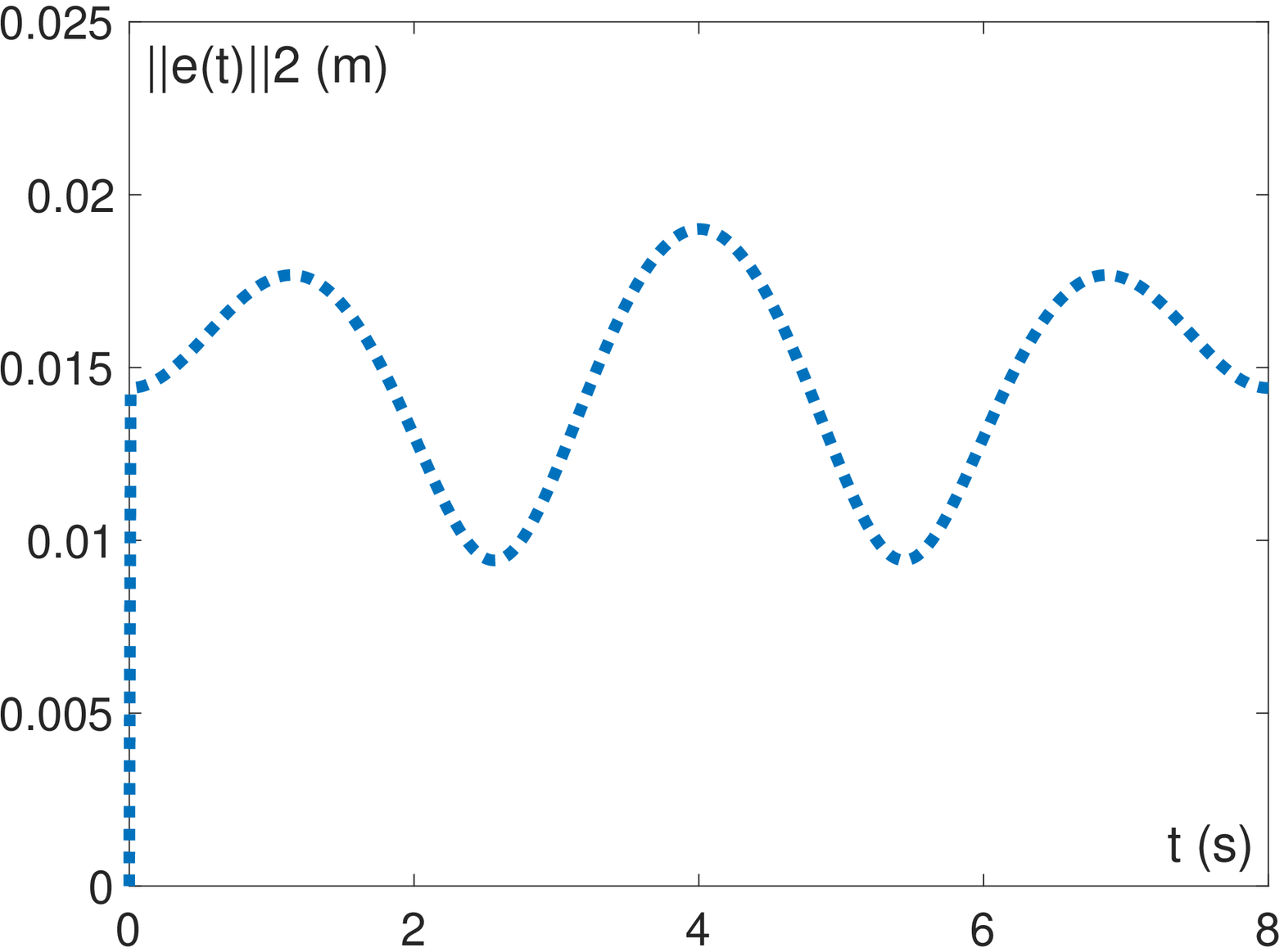}\label{fig.RN-5}}\\
  \subfigure []
  {\includegraphics[width=0.195\textwidth]{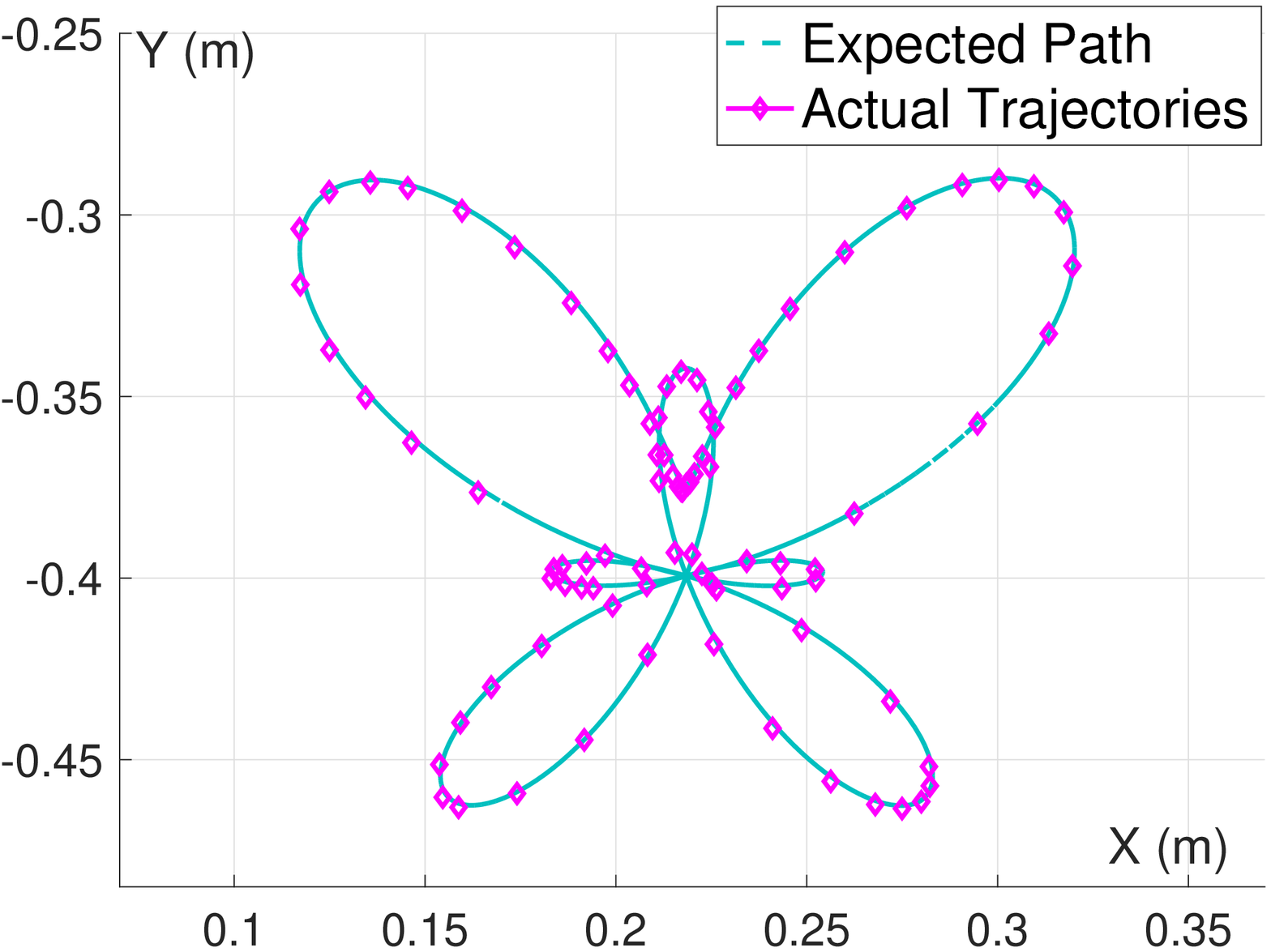}\label{fig.RM-6}}
  \subfigure []
  {\includegraphics[width=0.195\textwidth]{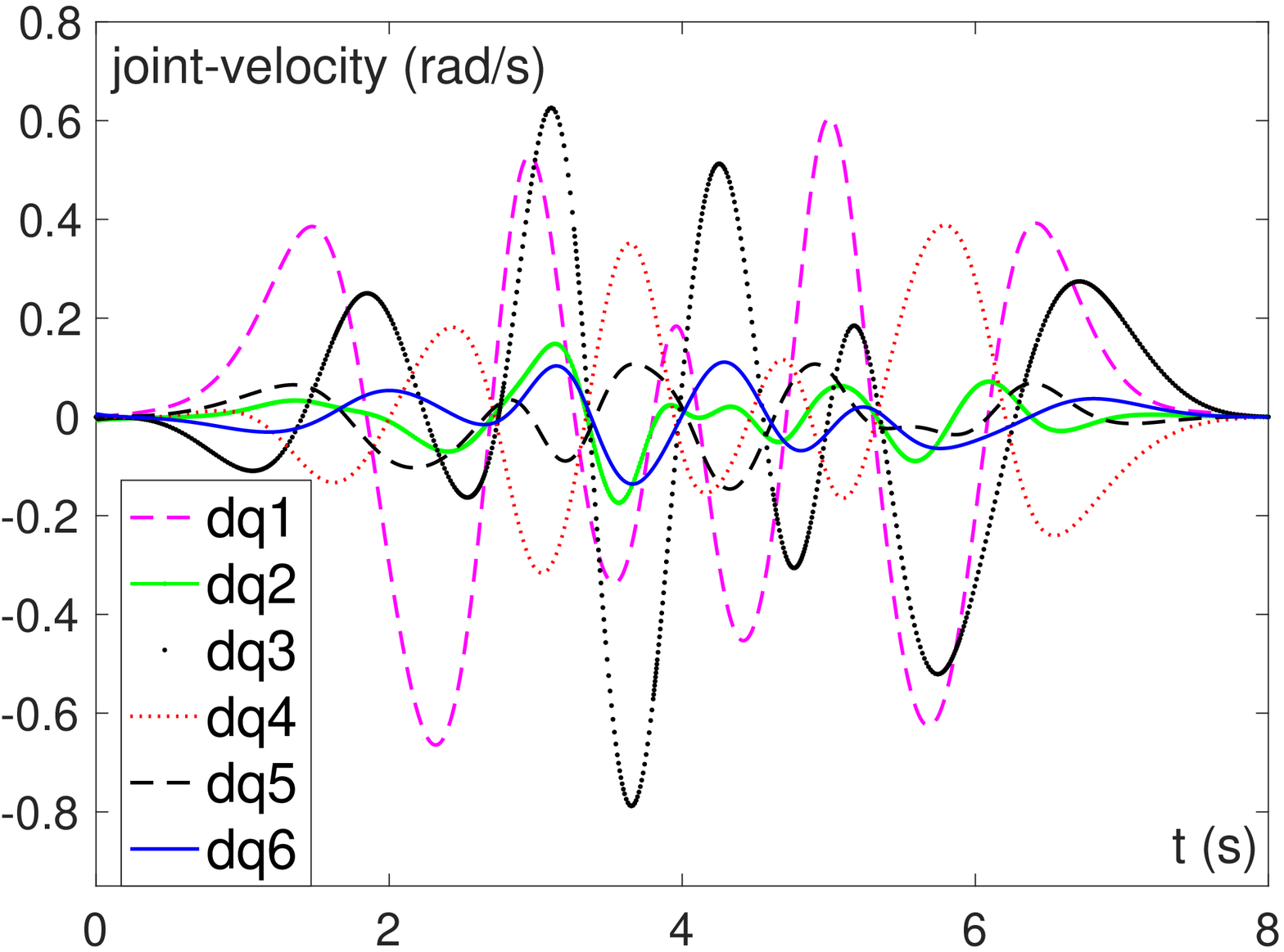}\label{fig.RM-7}}
  \subfigure []
  {\includegraphics[width=0.195\textwidth]{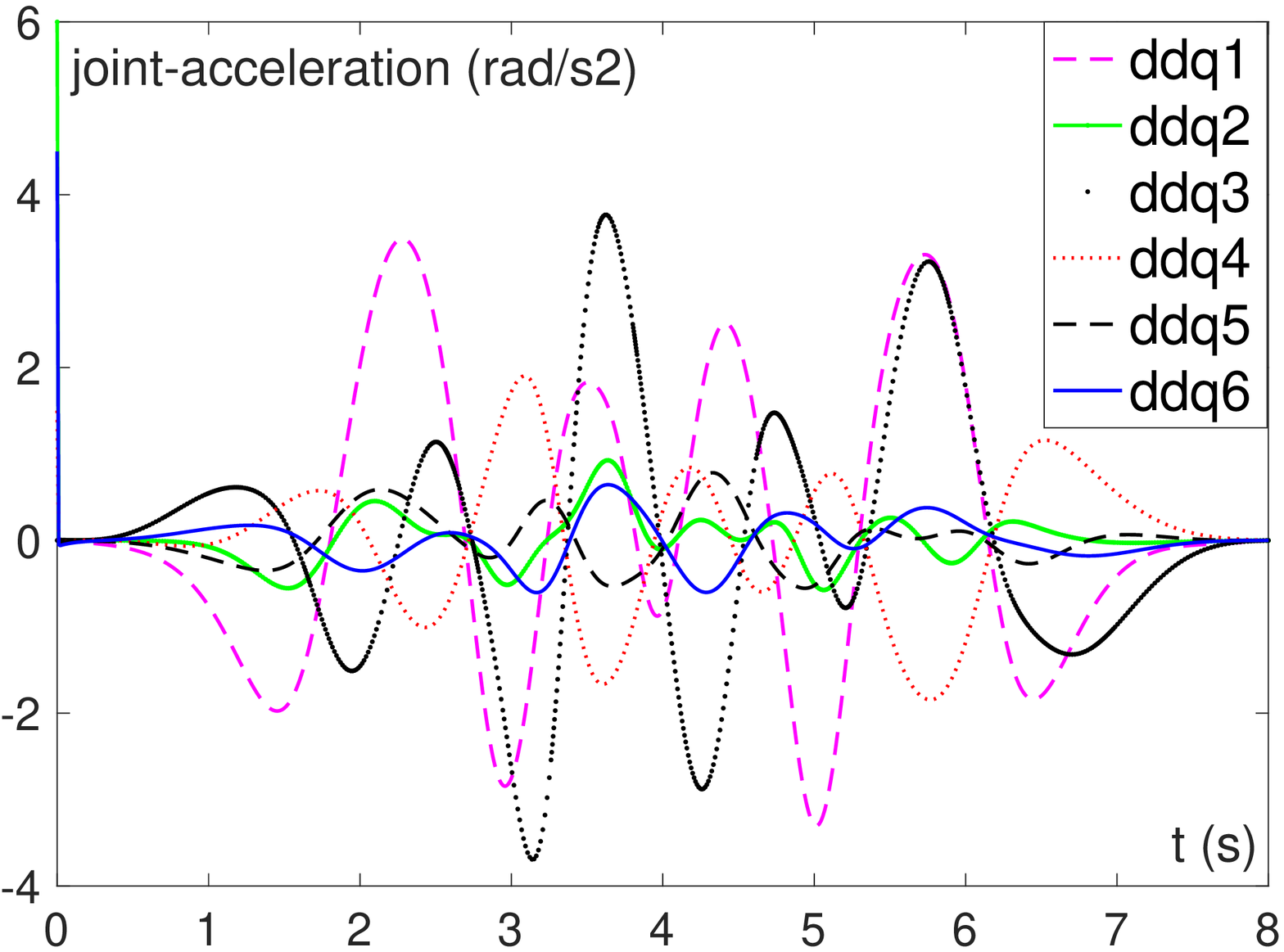}\label{fig.RM-8}}
  \subfigure []
  {\includegraphics[width=0.195\textwidth]{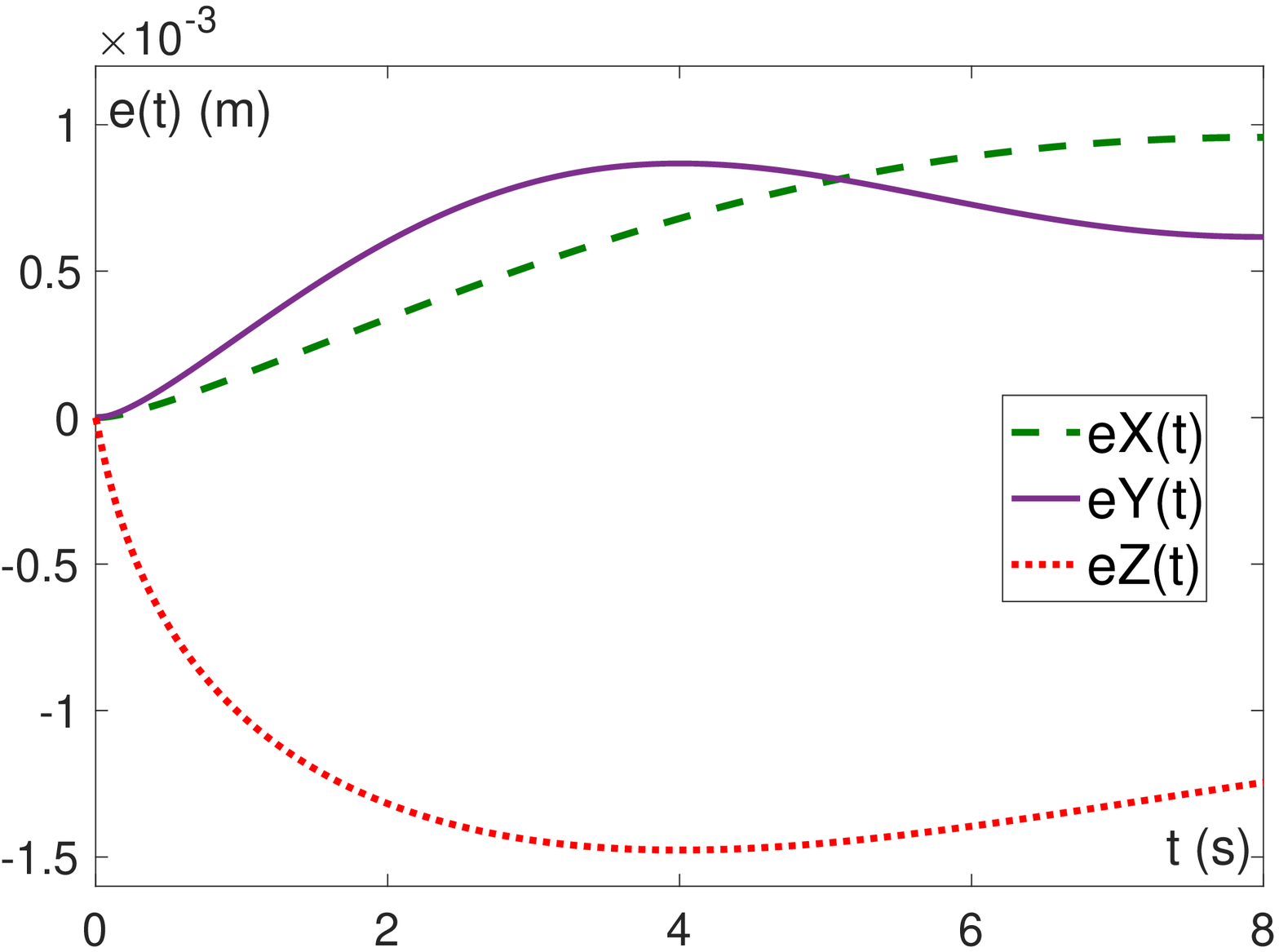}\label{fig.RM-9}}
  \subfigure []
  {\includegraphics[width=0.195\textwidth]{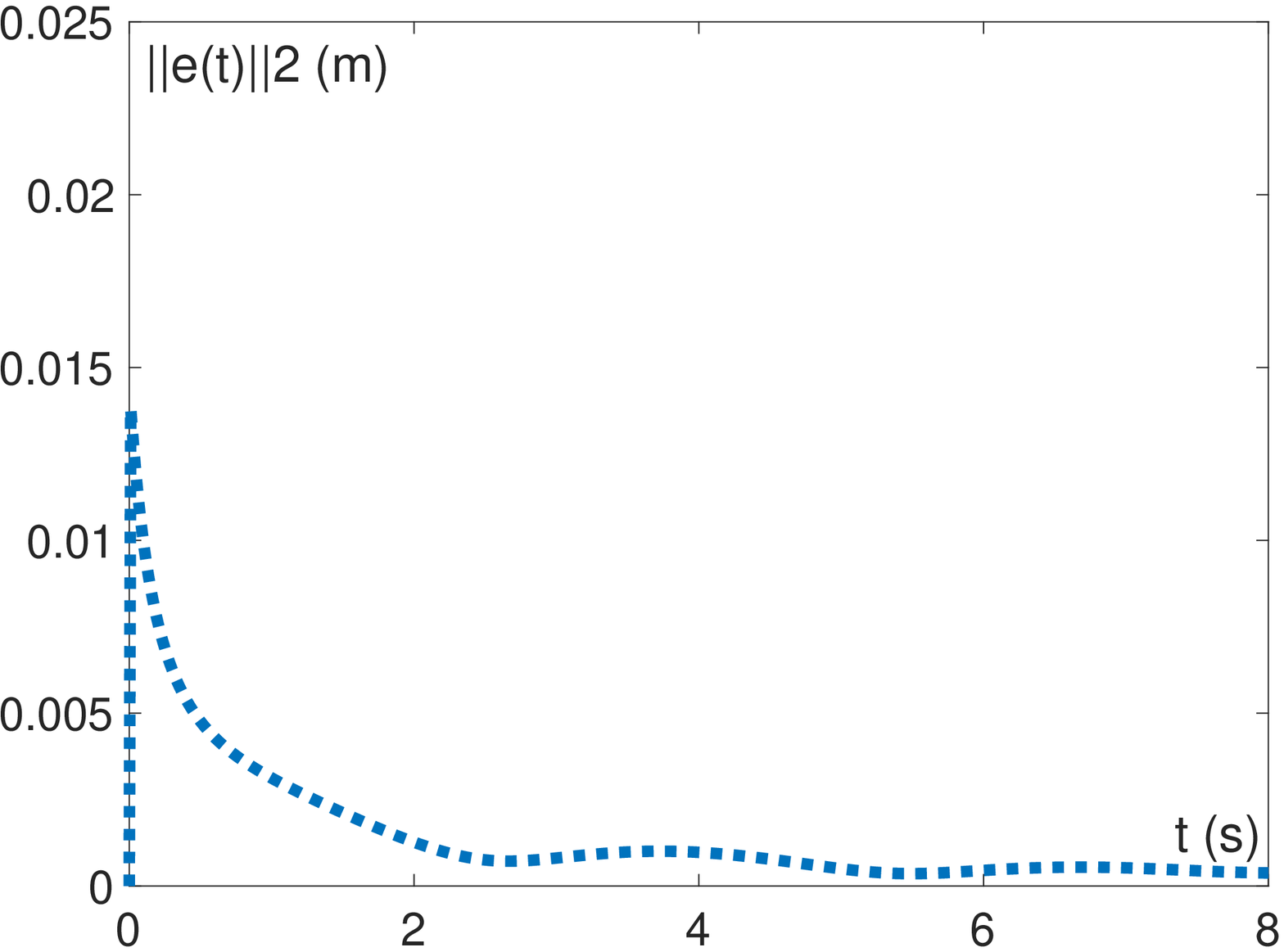}\label{fig.RM-10}}
  \caption{Computer simulation results via Hybrid Torque Scheme when tracking a butterfly path. (a) Trajectories of Z-RNN. (b) Joint-velocity of Z-RNN. (c) Joint-acceleration of Z-RNN. (d) Position error of Z-RNN. (e) RMS error of IE-RNN. (f) Trajectories of IE-RNN. (g) Joint-velocity of IE-RNN. (h) Joint-acceleration of IE-RNN. (i) Position error of IE-RNN. (j) RMS error of IE-RNN.}
  \label{fig.EXP2}
\end{figure*}

\subsection{Experiment of Repetitive Motion Scheme}

Based on the aforementioned Repetitive Motion Scheme (\ref{eqn.QP-RM}), comparative simulations between IE-RNN (\ref{eqn.IE-RNN}) and Z-RNN (\ref{eqn.Z-RNN}) are presented. The initial joint state $\theta(0)$ is set as $\theta(0)=[1.675, 2.843, -3.216, 4.187, -1.710, -2.650]^{\text{T}}$ (rad). The task execution time $T=8s$. The parameters of IE-RNN (\ref{eqn.IE-RNN}) and Z-RNN (\ref{eqn.Z-RNN}) are set as $\nu_1=500$ and $\nu_2=2500$. Without loss of generality, the time-varying noise is composed of a series of sine and cosine functions. Specifically, noise $\Delta\mathcal{N}_{2}(t)=[3\sin(t/\pi T), 6\cos(2t/\pi T), -7.5\sin(3t/\pi T), 1.5\cos(3t/\pi T),$ $-1.5\sin(3t/\pi T), 4.5\cos(t/\pi T), -1.5\sin(t\pi T), -1.5\sin(2t/$ $\pi T), 1.5\cos(t/\pi T)]^\text{T}$.

Firstly, a "starfish" shape path is introduced to verify the resolution of inverse kinematics of the proposed method. On the basis of the Repetitive Motion Scheme (\ref{eqn.QP-RM}), the simulation is synthesized by the proposed IE-RNN (\ref{eqn.IE-RNN}) and the traditional Z-RNN (\ref{eqn.Z-RNN}). And the simulative results include the desired path and actual trajectories, the joint-velocities $\dot{\theta}(t)$, the joint accelerations $\ddot{\theta}(t)$, the position errors $\epsilon_x$, $\epsilon_y$, $\epsilon_z$, and the root-mean-square (RMS) error, which is defined as $\sqrt{\epsilon_x^2+\epsilon_y^2+\epsilon_z^2}$.

The results of this simulative experiment are shown in Fig. \ref{fig.EXP1}. Specifically, sub-Figs. \ref{fig.EXP1}(a)-(e) denote the results of Z-RNN (\ref{eqn.Z-RNN}), and sub-Figs. \ref{fig.EXP1}(f)-(j) denote the results of IE-RNN (\ref{eqn.IE-RNN}). Evidently, with the disturbance of time-varying noises, the tracking trajectories mismatch the desired path when applying the traditional Z-RNN (\ref{eqn.Z-RNN}). On the contrary, the trajectories of IE-RNN (\ref{eqn.IE-RNN}) well matches the expected path. This can be quantitatively verified by the position errors and RMS errors. The maximal error generated by Z-RNN (\ref{eqn.Z-RNN}), which is visualized as $0.015m$, is about ten times larger than that of IE-RNN (\ref{eqn.IE-RNN}). What is more, the RMS error of Z-RNN (\ref{eqn.Z-RNN}) is divergent during the whole execution time, while the RMS error of IE-RNN (\ref{eqn.IE-RNN}) tends to zero.

\begin{figure*}
  \centering
  \includegraphics[width=0.99\textwidth]{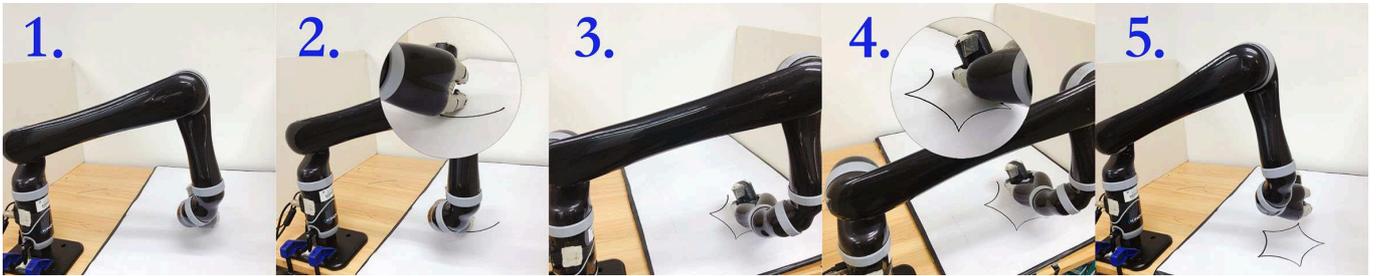}\\
  \caption{Practical experiment results via IE-RNN for tracking a starfish path on a Kinova JACO$^{2}$ robot manipulator.}
  \label{fig.EXPP1}
\end{figure*}

Secondly, a robot experiment is presented to verify the practical feasibility of the proposed IE-RNN (\ref{eqn.IE-RNN}). In terms of the computer simulations, a Kinova JACO$^{2}$ robot manipulator is utilized to the physical verification, and the snapshots are shown in Fig. \ref{fig.EXPP1}. As we can see, the end-effector of the manipulator, which holds a marker pen, is successfully tracks the desired "starfish" path.

\subsection{Experiment of Hybrid Torque Scheme}

On the basis of the Hybrid Torque Scheme (\ref{eqn.QP-HT}), a complex ''butterfly'' path is expected to be tracked by the IE-RNN (\ref{eqn.IE-RNN}) and Z-RNN (\ref{eqn.Z-RNN}) methods. The initial joint state $\theta(0)$ of the manipulator is set as $\theta(0)=[1.670, 2.845, -3.218, 4.182, -1.715, -2.655]^{\text{T}}$ (rad). Moreover, the task execution time $T$, parameters $\nu_1$ and $\nu_2$, and time-varying noise $\Delta\mathcal{N}_{2}(t)$ are set just the same as Repetitive Motion Scheme (\ref{eqn.QP-RM}). Similarly, the simulation results contain the matching degree of desired path and actual trajectories, the joint-velocities $\dot{\theta}(t)$, the joint accelerations $\ddot{\theta}(t)$, the position errors, and the RMS errors.

The simulative results are shown in Fig. \ref{fig.EXP2}. By comparing sub-Fig. \ref{fig.EXP2}(a) and sub-Fig. \ref{fig.EXP2}(f) we can draw the conclusion that the robustness performance of IE-RNN (\ref{eqn.IE-RNN}) is much better than that of Z-RNN (\ref{eqn.Z-RNN}). The trajectories synthesized by Z-RNN (\ref{eqn.Z-RNN}) mismatch the desired path again because the RMS error (see sub-Fig. \ref{fig.EXP2}(e)) is residuals in a relatively large value, i.e., about $0.024m$. On the contrary, the complex path tracking task is well finished by utilizing the proposed IE-RNN (\ref{eqn.IE-RNN}), which verifies the aforementioned theoretical analysis. The maximal RMS error of IE-RNN (\ref{eqn.IE-RNN}) is less than $0.015m$, and as the task goes on, the error decreases and finally approaches zero. It is worth pointing out that the joint angles generated by both IE-RNN (\ref{eqn.IE-RNN}) and Z-RNN (\ref{eqn.Z-RNN}) are return to the initial values, which can be verified by sub-Fig. \ref{fig.EXP2}(b) and sub-Fig. \ref{fig.EXP2}(g). In other words, the redundant kinematic resolution problem of joint drift can be modified by utilizing these two neural network methods.

\section{Concluding Remark and Future Work} \label{sec.Conclusion}

In this paper, an integration-enhanced recurrent neural network (IE-RNN) is proposed to obtain the kinematics resolution of redundant robot manipulator. Firstly, two schemes (i.e., Repetitive Motion Scheme and Hybrid Torque Scheme) are introduced and combined as a QP framework. Secondly, the proposed IE-RNN and the traditional Z-RNN are utilized to solve this QP problem. Compared with the traditional Z-RNN method, IE-RNN performs excellent feature when facing noise disturbance. MATLAB simulations and physical experiments further verify the feasibility, effectiveness, and accuracy of the proposed method.

Inspired by the ideas in Refs. \cite{ZhangA2017,8894424,9127132}, our future work is to explore a varying-parameter version for the IE-RNN model, and theoretically analyze the convergence and robustness.

\bibliographystyle{ieeetr}
\bibliography{NN}

\begin{thebibliography}{10}

\bibitem{Antonelli2009Stability}
G.~Antonelli, ``Stability analysis for prioritized closed-loop inverse
  kinematic algorithms for redundant robotic systems,'' {\em IEEE Transactions
  on Robotics}, vol.~25, no.~5, pp.~985--994, 2009.

\bibitem{NYR2018A}
Z.~Zhang, Y.~Niu, S.~Wu, S.~Lin, and L.~Kong, ``Analysis of influencing factors
  on humanoid robots’ emotion expressions by body language,'' in {\em
  Advances in Neural Networks – ISNN 2018, Lecture Notes in Computer Science,
  Springer, Cham}, vol.~10878, pp.~757--767, 2018.

\bibitem{zhang2020modification}
Z.~Zhang, L.~Kong, Y.~Niu, and Z.~Liang, ``Modification of
  gesture-determined-dynamic function with consideration of margins for motion
  planning of humanoid robots,'' {\em arXiv preprint arXiv:2008.06899},
  pp.~1--17, 2020.

\bibitem{Craig1986Introduction}
J.~J. Craig, {\em Introduction to Robotics: Mechanics and Control}.
\newblock Addison-Wesley Publishing Company, 1986.

\bibitem{Abdel2012A}
T.~M. Abdel-Rahman, ``A generalized practical method for analytic solution of
  the constrained inverse kinematics problem of redundant manipulators,'' {\em
  International Journal of Robotics Research}, vol.~10, no.~4, pp.~382--395,
  2012.

\bibitem{Shimizu2008Analytical}
M.~Shimizu, H.~Kakuya, W.~K. Yoon, K.~Kitagaki, and K.~Kosuge, ``Analytical
  inverse kinematic computation for 7-dof redundant manipulators with joint
  limits and its application to redundancy resolution,'' {\em IEEE Transactions
  on Robotics}, vol.~24, no.~5, pp.~1131--1142, 2008.

\bibitem{Merat2003Introduction}
F.~Merat, ``Introduction to robotics: Mechanics and control,'' {\em IEEE
  Journal on Robotics \& Automation}, vol.~3, no.~2, pp.~166--166, 2003.

\bibitem{Angeles1985On}
J.~Angeles, ``On the numerical solution of the inverse kinematic problem,''
  {\em International Journal of Robotics Research}, vol.~4, no.~2, pp.~21--37,
  1985.

\bibitem{Chevallereau1987Efficient}
C.~Chevallereau and W.~Khalil, ``Efficient method for the calculation of the
  pseudo inverse kinematic problem,'' in {\em IEEE International Conference on
  Robotics and Automation. Proceedings}, pp.~1842--1848, 1987.

\bibitem{Toshani2014Real}
H.~Toshani and M.~Farrokhi, ``Real-time inverse kinematics of redundant
  manipulators using neural networks and quadratic programming: A
  lyapunov-based approach,'' {\em Robotics and Autonomous Systems}, vol.~62,
  no.~6, pp.~766--781, 2014.

\bibitem{Zhang2013Variable}
Z.~Zhang and Y.~Zhang, ``Variable joint-velocity limits of redundant robot
  manipulators handled by quadratic programming,'' {\em IEEE/ASME Transactions
  on Mechatronics}, vol.~18, no.~2, pp.~674--686, 2013.

\bibitem{He1994Solving}
B.~He, ``Solving a class of linear projection equations,'' {\em Numerische
  Mathematik}, vol.~68, no.~1, pp.~71--80, 1994.

\bibitem{KLD2018A}
Z.~Zhang, L.~Kong, and Y.~Niu, ``A time-varying-constrained motion generation
  scheme for humanoid robot arms,'' in {\em Advances in Neural Networks –
  ISNN 2018, Lecture Notes in Computer Science, Springer, Cham}, vol.~10878,
  pp.~775--785, 2018.

\bibitem{Zhang2017Three}
Z.~Zhang, L.~Zheng, J.~Yu, Y.~Li, and Z.~Yu, ``Three recurrent neural networks
  and three numerical methods for solving a repetitive motion planning scheme
  of redundant robot manipulators,'' {\em IEEE/ASME Transactions on
  Mechatronics}, vol.~22, no.~3, pp.~1423--1434, 2017.

\bibitem{8665072}
Z.~{Zhang}, L.~{Kong}, Z.~{Yan}, K.~{Chen}, S.~{Li}, X.~{Qu}, and N.~{Tan},
  ``Comparisons among six numerical methods for solving repetitive motion
  planning of redundant robot manipulators*,'' in {\em 2018 IEEE International
  Conference on Robotics and Biomimetics (ROBIO)}, pp.~1645--1652, 2018.

\bibitem{Tejomurtula1999Inverse}
S.~Tejomurtula and S.~Kak, ``Inverse kinematics in robotics using neural
  networks,'' {\em Information Sciences}, vol.~116, pp.~147--164, 1999.

\bibitem{Xia2000A}
Y.~Xia and J.~Wang, {\em A recurrent neural network for solving linear
  projection equations}.
\newblock Elsevier Science Ltd., 2000.

\bibitem{Mitra2015Cerebellum}
A.~E. Mitra, E.~M. Mehdi, S.~H. Mehran, D.~Christian, and A.~O.~N. Azuan,
  ``Cerebellum-inspired neural network solution of the inverse kinematics
  problem,'' {\em Biological Cybernetics}, vol.~109, no.~6, pp.~561--574, 2015.

\bibitem{Qu2009Real}
H.~Qu, S.~X. Yang, A.~R. Willms, and Z.~Yi, ``Real-time robot path planning
  based on a modified pulse-coupled neural network model,'' {\em IEEE
  Transactions on Neural Networks}, vol.~20, no.~11, pp.~1724--1739, 2009.

\bibitem{Zhang2009Zhang}
Y.~Zhang and Z.~Li, ``Zhang neural network for online solution of time-varying
  convex quadratic program subject to time-varying linear-equality
  constraints,'' {\em Physics Letters A}, vol.~373, pp.~1639--1643, 2009.

\bibitem{9097909}
Z.~{Hu}, L.~{Xiao}, J.~{Dai}, Y.~{Xu}, Q.~{Zuo}, and C.~{Liu}, ``A unified
  predefined-time convergent and robust znn model for constrained quadratic
  programming,'' {\em IEEE Transactions on Industrial Informatics}, vol.~PP,
  pp.~1--1, 2020.

\bibitem{Zhang2010Robustness}
Z.~{Zhang}, L.~{Kong}, L.~{Zheng}, P.~{Zhang}, X.~{Qu}, B.~{Liao}, and Z.~{Yu},
  ``Robustness analysis of a power-type varying-parameter recurrent neural
  network for solving time-varying qm and qp problems and applications,'' {\em
  IEEE Transactions on Systems, Man, and Cybernetics: Systems}, vol.~PP,
  pp.~1--14, 2018.

\bibitem{Zhang2008The}
Y.~Zhang, W.~Ma, and C.~Yi, ``The link between newton iteration for matrix
  inversion and zhang neural network (znn),'' in {\em IEEE International
  Conference on Industrial Technology}, pp.~1--6, 2008.

\bibitem{Guo2014Li}
D.~Guo and Y.~Zhang, ``Li-function activated znn with finite-time convergence
  applied to redundant-manipulator kinematic control via time-varying jacobian
  matrix pseudoinversion,'' {\em Applied Soft Computing Journal}, vol.~24,
  pp.~158--168, 2014.

\bibitem{9072323}
Z.~{Zhang}, L.~{Zheng}, Z.~{Chen}, L.~{Kong}, and H.~R. {Karimi},
  ``Mutual-collision-avoidance scheme synthesized by neural networks for dual
  redundant robot manipulators executing cooperative tasks,'' {\em IEEE
  Transactions on Neural Networks and Learning Systems}, vol.~PP, pp.~1--15,
  2020.

\bibitem{8877851}
Z.~{Zhang}, X.~{Deng}, L.~{Kong}, and S.~{Li}, ``A circadian rhythms learning
  network for resisting cognitive periodic noises of time-varying dynamic
  system and applications to robots,'' {\em IEEE Transactions on Cognitive and
  Developmental Systems}, vol.~PP, pp.~1--1, 2019.

\bibitem{8558699}
Z.~{Zhang}, X.~{Deng}, X.~{Qu}, B.~{Liao}, L.~{Kong}, and L.~{Li}, ``A
  varying-gain recurrent neural network and its application to solving online
  time-varying matrix equation,'' {\em IEEE Access}, vol.~6, pp.~77940--77952,
  2018.

\bibitem{ZhangA2017}
Z.~{Zhang}, L.-D. {Kong}, and L.~{Zheng}, ``Power-type varying-parameter rnn
  for solving tvqp problems: Design, analysis, and applications,'' {\em IEEE
  Transactions on Neural Networks and Learning Systems}, vol.~30, no.~8,
  pp.~2419--2433, 2019.

\bibitem{Mazenc2006Further}
F.~Mazenc and M.~Malisoff, ``Further constructions of control-lyapunov
  functions and stabilizing feedbacks for systems satisfying the
  jurdjevic-quinn conditions,'' {\em IEEE Transactions on Automatic Control},
  vol.~51, no.~2, pp.~360--365, 2006.

\bibitem{Luse1988A}
D.~W. Luse, ``A nyquist-type stability test for multivariable distributed
  systems,'' {\em IEEE Transactions on Automatic Control}, vol.~33, no.~6,
  pp.~563--566, 1988.

\bibitem{Pagilia2001Control}
P.~R. Pagilia, ``Control of contact problem in constrained euler-lagrange
  systems,'' {\em IEEE Transactions on Automatic Control}, vol.~46, no.~10,
  pp.~1595--1599, 2001.

\bibitem{8894424}
Y.~{Qi}, L.~{Jin}, Y.~{Wang}, L.~{Xiao}, and J.~{Zhang}, ``Complex-valued
  discrete-time neural dynamics for perturbed time-dependent complex quadratic
  programming with applications,'' {\em IEEE Transactions on Neural Networks
  and Learning Systems}, vol.~PP, pp.~1--15, 2019.

\bibitem{9127132}
L.~{Jia}, L.~{Xiao}, J.~{Dai}, and Y.~{Cao}, ``A novel fuzzy-power zeroing
  neural network model for time-variant matrix moore-penrose inversion with
  guaranteed performance,'' {\em IEEE Transactions on Fuzzy Systems}, vol.~PP,
  pp.~1--1, 2020.

\end{thebibliography}

\end{document}